\documentclass[9pt,twocolumn,twoside]{article}
\usepackage[top=3cm,bottom=3cm,right=2cm,left=2cm]{geometry}
\usepackage{authblk}
\usepackage{graphicx}
\usepackage{verbatim}
\usepackage{amsmath}
\usepackage{amssymb}
\usepackage{mathrsfs}
\usepackage{cite}
\usepackage{hyperref}
\usepackage[title]{appendix}
\usepackage{subcaption}

\newcommand{\bea}{\begin{eqnarray}}
\newcommand{\eea}{\end{eqnarray}}

\newcommand{\be}{\begin{equation}}
\newcommand{\ee}{\end{equation}}
\newcommand{\ba}{\begin{eqnarray}}
\newcommand{\ea}{\end{eqnarray}}

\newcommand{\NA}{N_{\rm A}}
\newcommand{\NC}{N_{\rm C}}
\newcommand{\NAi}{\tilde N_{\tilde{\rm A}}}

\begin{document}
\title{Emergence of homochirality in large molecular systems}
\author[a,1]{Gabin Laurent}
\author[a]{David Lacoste}
\author[b]{Pierre Gaspard}

\affil[a]{Gulliver, UMR CNRS 7083, \'Ecole Supérieure de Physique et de Chimie Industrielles de la Ville de Paris, Paris Sciences et Lettres Research university, F-75231 Paris, France}
\affil[b]{Center for Nonlinear Phenomena and Complex Systems, Université Libre de Bruxelles, B-1050 Brussels, Blegium.}
\affil[1]{gabin.laurent@espci.fr}
\date{}

\twocolumn[
\begin{@twocolumnfalse}
\maketitle
\begin{abstract}
The selection of a single molecular handedness, or homochirality across all living matter, is a mystery in the origin of life. 
Frank's seminal model showed in the fifties how chiral symmetry breaking can occur in non-equilibrium chemical networks.
However, an important shortcoming in this classic model is that it considers a small number of species, while there is no reason for the prebiotic system, in which homochirality first appeared, to have had such a simple composition. Furthermore, this model does not provide information on what could have been the size of the molecules involved in this homochiral prebiotic system.
Here, we show that large molecular systems are likely to undergo a phase transition towards a homochiral state, as a consequence of the fact that they contain a large number of chiral species. 
Using chemoinformatics tools, we quantify how abundant are chiral species in the chemical universe of all possible molecules of a given length.
Then, we propose that Frank's model should be extended to include a large number of species, in order to possess the transition towards homochirality as confirmed by numerical simulations.   
Finally, using random matrix theory, we prove that large non-equilibrium reaction networks possess a generic and robust phase transition towards a homochiral state.
\end{abstract}
\vspace{0.5cm}

\textbf{Keywords:} homochirality, origin of life, prebiotic chemistry, random matrices, statistical physics \vspace{1cm}
\end{@twocolumnfalse}
]



Life on Earth relies on chiral molecules$-$that is, species not superposable on their mirror
images. 
A given biological molecule forms with its mirror image a pair of enantiomers. Homochirality precisely means the dominance of one member of the pair across the entire biosphere.
For instance in our cells, 
biochemical reaction networks only involve left-handed (L-chiral) amino-acids and right-handed (D-chiral) sugars,
but the reason for this absolute specificity escapes us and is one of the most fascinating questions in the origin of life. 

The origin of homochirality comes with two questions and related observations: what caused the initial biais of 
one enantiomer over the other in the presumably racemic environment
of the prebiotic world and how was this bias sustained and maintained as in today's biological world \cite{blackmond_origin_2010} ?
It is believed that mineral surfaces on earth \cite{E13} or circularly polarized light in interstellar space \cite{meinert_photochirogenesis_2011} could explain the first observation, while models based on non-equilibrium reaction networks can explain the second observation \cite{F53,KN83,PKBCA07,KA01,WC05,KML10,SH13,SRBH16,HBAR17,jafarpour_noise-induced_2017,JBG15,PBC04}. 

However, there is an important shortcoming in the common discussions addressing the issue of homochirality$-$namely that they only consider a small number of chiral species, as in Frank's classic model ~\cite{F53}, 
or in its first experimental realization more than forty years later by Soai et al. \cite{soai_asymmetric_1995}. There is no reason to expect that the prebiotic world, in which homochirality first emerged, had such a simple and homogeneous chemical composition. Instead, it is more natural to assume that this composition was complex, heterogeneous and included a large number of chiral and achiral species. 
We show in this paper that generic non-equilibrium reaction networks 
possess a phase transition towards a homochiral state as a consequence of the fact that the number of chiral species becomes large. 

\subsection*{Cross-Over between chiral and achiral chemical worlds}

\begin{figure*}[ht]
\centering
\includegraphics[width=1\linewidth]{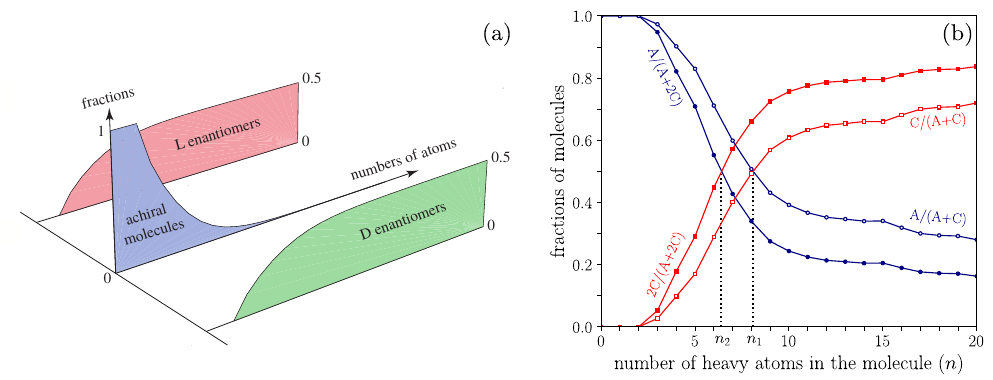}
\caption{Cross-over between the achiral and chiral chemical worlds. (a) A three-dimensional schematic representation of the fractions of possible achiral and chiral molecules as function of their number of atoms. Every chiral molecule appears as mirror-reflected D- and L-enantiomers. (b) A two-dimensional representation of the fractions of achiral (circles) and chiral (squares) molecules in an expanded PubChem database containing all the stereoisomers of molecules with at most $n \leq 20$ heavy atoms. Filled symbols correspond to counting stereoisomers twice, while empty symbols correspond to counting them once in the fractions. Error bars are smaller than symbol sizes and the cross-over occurs at $n_2 \simeq 6.4$ and $n_1 \simeq 8.1$ depending on which counting is considered (SI Appendix, section \ref{Main}).}
\label{fig1}
\end{figure*}

With this aim, we first ask how abundant are chiral species in the chemical universe of all possible molecules?
It turns out that chirality is rare among molecules with a small number of atoms, but that possible chiral stereoisomers multiply as their number of atoms increases.  Accordingly, we should expect a cross-over between the achiral world of small molecules and the chiral world of large molecules involved in chemical reaction networks.
The cross-over should be characterized by some specific number of atoms where the fractions of achiral and chiral molecules become equal, as schematically depicted in (Fig.~\ref{fig1}a).  
Beyond the cross-over, the chiral molecules are dominant over achiral ones. 
The issue of this cross-over is 
important, in particular, because it associates the emergence of homochirality with some molecular size.
Starting with monosubstituted alkanes and alkanes for which an exact enumeration of 
stereoisomers is available \cite{fujita_alkanes_2007, F07}, we find that the cross-over measured in number of carbon atoms in these molecules, lies between $4.7$ and $5.7$ for monosubstituted alkanes, and between $8.4$ and $9.5$ for alkanes.

In the chemical universe of Fink and Reymond \cite{FR07}, which contains all virtual molecules of a given number of heavy atoms ({\it i.e.} atoms heavier than hydrogen) and satisfying some basic set of rules of chemistry, we find a cross-over at $8.5$ heavy atoms. Since a similar study was not available for real molecules, we turn to the chemical database PubChem \cite{pubchem_database}. From the raw data of this large database, we find that a cross-over occurs for molecules of $9.4$ heavy atoms (SI Appendix, section \ref{Main}). However, since many chiral molecules do not have all their enantiomers or stereoisomers, we have also analyzed an expanded database, in which every chiral molecule contains either all its enantiomers or all its stereoisomers. In Fig.~\ref{fig1}b, the fractions of chiral and achiral species are shown for the case of the database expanded in stereoisomers.   
Our results for the various estimates for the cross-over are gathered in Table~\ref{table1} and extracted from Fig.~\ref{fig1}b and SI Appendix, Figs.~\ref{fig2}, \ref{fig3}, \ref{fig:raw_pubchem} and \ref{fig:enantiomer_pubchem}. Remarkably, irrespective of the precise procedure to generate and analyze the database and regardless of the precise composition of the molecules, the cross-over between the achiral and chiral worlds occurs for a number of heavy atoms of the order of about $10$.  
The main consequence of this cross-over is that the stereoisomer distribution goes from unimodal (with a maximum for achiral molecules) to bimodal (with maxima for opposite enantiomers) as the length of molecules increases. This emerging bimodality is potentially susceptible to induce a chiral symmetry breaking.

\begin{table}[h]
\centering
\caption{Positions of the cross-over $n_1$ and $n_2$, measured in terms of number of carbon atoms, in the study of alkanes and monosubstituted alkanes, or in terms of number of heavy atoms in the other studies. The estimate $n_1$ (resp. $n_2$) is obtained by counting once (resp. twice) the pairs of enantiomers.}
\begin{tabular}{lcc}
Data & $n_1$ & $n_2$ \\
\hline
monosubstituted alkanes stereoisomers & 5.7 & 4.7 \\
alkanes stereoisomers & 9.5 & 8.4 \\
Chemical Universe & 8.5 & $-$ \\
PubChem database using raw data & 9.4* & $-$ \\
PubChem database using generated & & \\ enantiomers & 12.7 & 6.7 \\
PubChem database using generated & & \\ stereoisomers & 8.1 & 6.4 \\
\hline
\end{tabular}

$^*$ The cross-over for PubChem raw data occurs between $n_1$ and $n_2$ because not all enantiomers of a given species are present in the database.
\label{table1}
\end{table}

\subsection*{Spontaneous symmetry breaking into a chiral state}

\begin{figure*}[ht]
\centering
\includegraphics[width=1.02\linewidth]{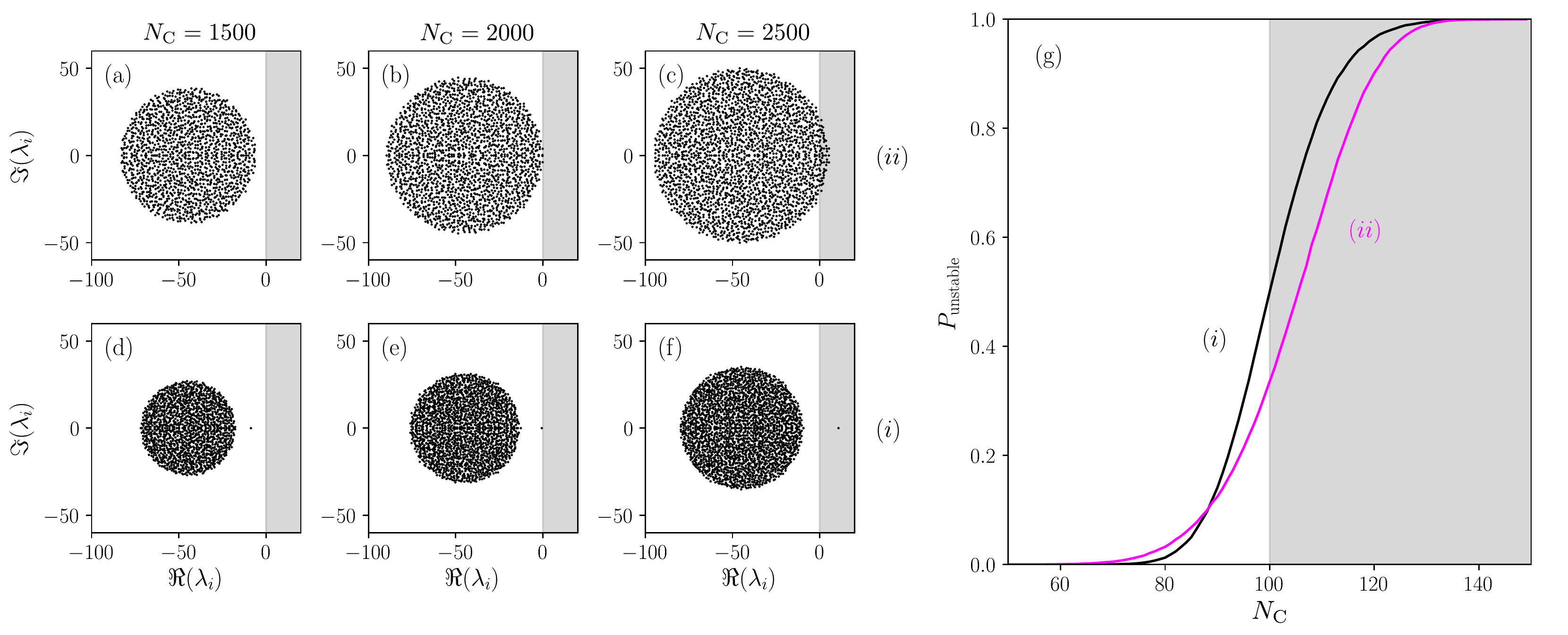}
\caption{Distribution of eigenvalues and corresponding instability criterion. Subfigures (a), (b) and (c) represent the eigenvalues of the random matrix ${\boldsymbol{\mathsf M}}$ with Gaussian entries with $\mu = 0$, $\sigma = 1$ and $\tau = 1/\sqrt{2000}$, where the instability mechanism $(ii)$ occurs as $N_{\rm C}$ increases, {\it i.e.} the edge of the Girko's circle densely filled by eigenvalues crosses the zero real axis. Subfigures (d), (e) and (f) depict the mechanism $(i)$ where the zero real axis is crossed by an isolated eigenvalue. In this case, the matrix ${\boldsymbol{\mathsf M}}$ is a random matrix with Gaussian entries of parameters $\mu = 1/\sqrt{2000}$, $\sigma = 0.7$ and $\tau = 1/\sqrt{2000}$. Grey colored area represents the positive real part zone of the complex plane, {\it i.e.} the area in which an eigenvalue causes the instability. Sizes of random matrices ${\boldsymbol{\mathsf M}}$ are $N_{\rm C}=1500$ for subfigures (a) and (d), $N_{\rm C}=2000$ for subfigures (b) and (e) and $N_{\rm C}=2500$ for subfigures (c) and (f). Subfigure (g) represents the probability for the racemic state of the chemical system to be unstable as a function of $N_{\rm C}$. 
Black curve represents the mechanism $(i)$ for a Gaussian random matrix ${\boldsymbol{\mathsf M}}$ with $\mu = 1/\sqrt{100}$, $\sigma = 0.7$ and $\tau = 1/\sqrt{100}$. Magenta curve represents the mechanism $(ii)$ for a Gaussian random matrix ${\boldsymbol{\mathsf M}}$ with $\mu = 0$, $\sigma = 1$ and $\tau = 1/\sqrt{100}$. Grey colored area represents the theoretical instability area, which is $N>100$ for both mechanisms here, given the choice of parameters $\mu$, $\sigma$ and $\tau$. The statistics has been carried out over $100,000$ realizations of random matrices.
Note also that the two curves in subfigure (g) for the two mechanisms, were drawn for different parameters $\mu$ and $\sigma$.}
\label{figGinibre}
\end{figure*}

We now come to our central point$-$namely, on how to explain the emergence of homochirality from the multiplication of chiral species in non-equilibrium reaction networks.  
Specifically, we consider a reaction network involving achiral and chiral species 
described by the concentration vector ${\bf c}$, which contains the vector ${\bf c}_{\rm D}$ (resp. 
${\bf c}_{\rm L}$) for the $N_{\rm C}$ D-enantiomers (resp. for the $N_{\rm C}$ L-enantiomers) and the vector ${\bf c}_{\rm A}$ for the remaining $N_{\rm A}$ achiral species.  
In such a system, the evolution of the concentrations is ruled by 
\be\label{kin_eqs}
\frac{d{\bf c}}{dt} = {\bf F}({\bf c}) + \frac{1}{\tau}({\bf c}_0-{\bf c}) \, , 
\ee
where ${\bf c_0}$ is the concentration vector of the species supplied from the environment at the rate $1/\tau$ and responsible for driving the system out of equilibrium and ${\bf F}({\bf c})={\boldsymbol\nu}\cdot{\bf w}({\bf c})$  are the reaction rates with specific chiral symmetry  (Eq.~(\ref{mirror_symm}) in Materials and Methods), which need not obey mass-action law. In this expression, ${\boldsymbol\nu}$ is the matrix of stoichiometric coefficients, ${\bf w}({\bf c})$ the set of net reaction rates. 
After reaction, the species in excess are flowing out of the system at the same rate $1/\tau$ as for the supply, so that $\tau$ represents the mean residence time of the species in the system.

The stability of these equations may be characterized 
by linearizing them about the racemic state, which is defined by the condition ${\bf c}_{\rm D}={\bf c}_{\rm L}$, and
is assumed to exist in a steady state. With the small parameter $\delta{\bf x}$, 
where ${\bf x}$ denotes the chiral enantiomeric excess ${\bf x} \equiv \frac{1}{2} ({\bf c}_{\rm L}-{\bf c}_{\rm D})$, we obtain

\be
\frac{d}{dt} \delta{\bf x} = \left( {\boldsymbol{\mathsf J}}-\frac{1}{\tau} \right) \cdot\delta{\bf x} + \frac{1}{\tau}\, \delta{\bf x}_0 \, ,
\ee
where ${\boldsymbol{\mathsf J}}$ represents the Jacobian matrix
deduced from the kinetic equations~(\ref{kin_eqs}). The racemic mixture is unstable if at least one of the eigenvalues of the matrix ${\boldsymbol{\mathsf M}}={\boldsymbol{\mathsf J}}-\frac{1}{\tau}\, {\boldsymbol{\mathsf I}}$ (with ${\boldsymbol{\mathsf I}}$ the identity matrix) has a positive real part. In a large reaction network, the reaction rates may take very different values, so that the matrix ${\boldsymbol{\mathsf J}}$ may be treated as a random matrix \cite{allesina_stability_2012,may_robert_will_1972,gardner_connectance_1970}. The simplest model is to assume that the elements of this matrix are independent and identically distributed real numbers (but not necessarily Gaussian distributed) of mean value $\mu$ and variance $\sigma^2$ \cite{ginibre_statistical_1965}.
When $\mu=0$, random matrix theory shows that the complex eigenvalues are uniformly distributed in a disk of radius $\sigma \sqrt{N_{\rm C}}$ in the limit of large values of $N_{\rm C}$ \cite{girko_circular_1985}. When $\mu \neq 0$, we find that there exists a single and isolated eigenvalue,
which is equal to $\mu N_{\rm C}$, and the corresponding eigenvector has uniform components to dominant order (SI Appendix, section \ref{Jacobian-matrix}). Two possible mechanisms for the instability of the racemic state then emerge for large $ N_{\rm C}$ (SI Appendix, section \ref{GIC}). Either (i) the instability occurs due to 
the isolated eigenvalue 
as illustrated in Fig.~\ref{figGinibre}d,e and~f; otherwise (ii) it occurs due to the eigenvalues located on the edge of the circle (which may be real or complex valued) as illustrated in Fig.~\ref{figGinibre}a,b and~c. It follows from this that when $\mu>0$ and $N_{\rm C}  \ge \max\{1/(\tau \mu), (\sigma/ \mu)^2 \}$, the system becomes unstable by the first mechanism where all species become simultaneously unstable, and when $(\sigma/\mu)^2 \ge N_{\rm C} \ge 1/(\tau \sigma)^2$, the system becomes unstable by the second mechanism and in this case only a subpart of all the species become unstable at the transition. In such cases, random matrix theory predicts that as $N_{\rm C}$ becomes large, these mechanisms of instability become more and more likely. This is confirmed by the shape of the probability for the racemic state to be unstable versus $N_{\rm C}$ shown in Fig.~\ref{figGinibre}g for both mechanisms. 
If the matrix elements are statistically correlated, the non-dominant eigenvalues may have a different distribution, but the isolated eigenvalue behaves similarly.

\subsection*{The effect of chiral species multiplication}

\begin{figure*}[ht]
\centering
\includegraphics[width=1\linewidth]{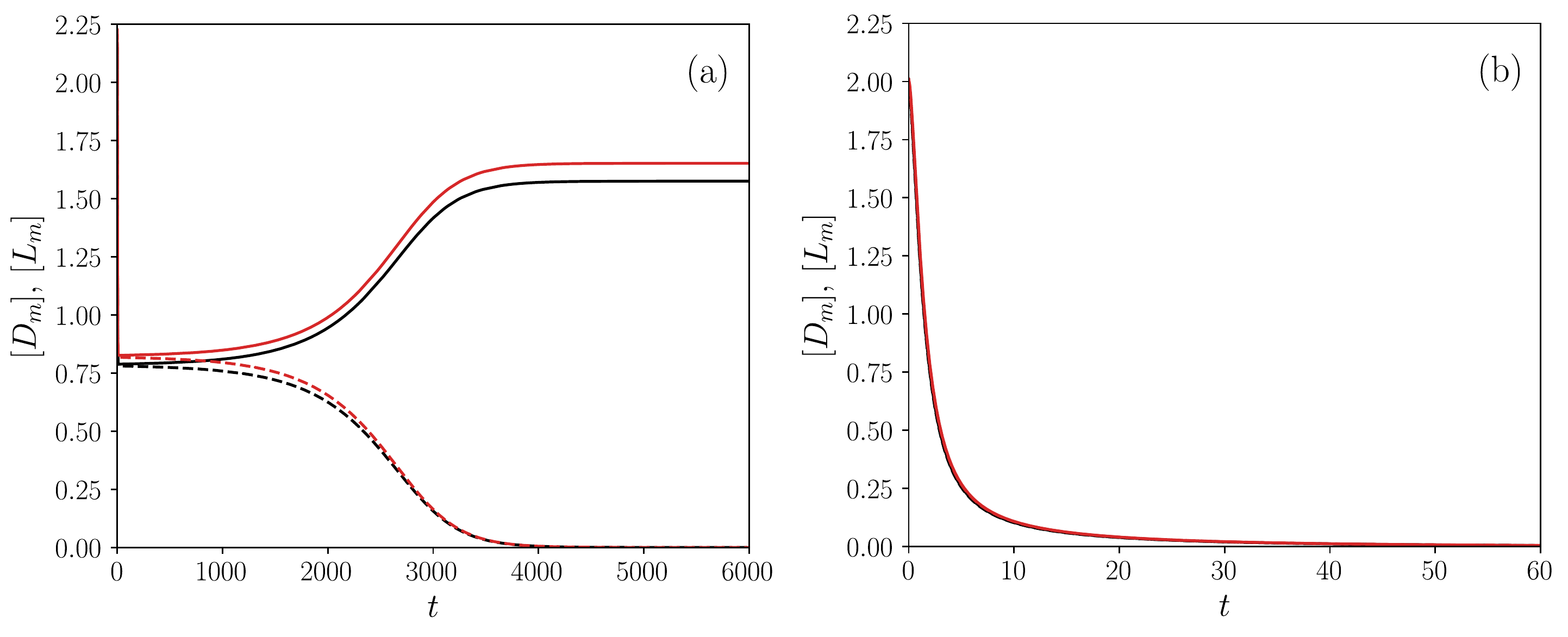}
\caption{Dynamical simulations of the autocatalytic network (\ref{react-ADDD-num})-(\ref{react-ALLL-num})-(\ref{react-AADL-num}). Typical time evolution of two species contained in the autocatalytic network as a function of time (a) above the threshold concentration $A_0$ and (b) below it.  The solid lines represent one of the two enantiomers for a given species and the dashed line, the other enantiomer. Both simulations were carried out with an initial enantiomeric excess $\epsilon = 10^{-2}$ and D- and L-enantiomers concentrations of all chiral species were initialized at $D_0 = 2 + \epsilon$ and $L_0 = 2-\epsilon$. The unactivated achiral specie was initialized at $\tilde A_0 = 0$ and the activated one at $A_0 = 80$ in (a) and $A_0 = 45$ in (b). All the constants $k_{+ijk}$ and $\tilde k_{-ij}$ follow a log-normal distribution of parameters $\mu = -10.02$ and $\sigma = 1.27$ (\textit{i.e.}, corresponding to a log-normal distribution with $\langle k_+ \rangle = \langle \tilde k_- \rangle = 10^{-4}$ and $\sigma_{k_+} = \sigma_{\tilde k_-} = 2\times 10^{-4}$), with $\tilde k_{ij} = \tilde k_{ji}$ to satisfy the mirror symmetry described in Eq.~(\ref{mirror_symm}). The number of chiral species was set up to $\NC = 20$.}
\label{fig-frank}
\end{figure*}

In order to show that this general scenario can be realized in practice in a non-equilibrium reaction network, 
we now introduce generalizations of Frank's model, in which we have multiplied the number of chiral and achiral species and we have assumed an arbitrary assignation L or D to each enantiomer. We also include reverse reactions in order to guarantee the compatibility with the existence of an equilibrium state even though the system is driven out of equilibrium. It is essential that the system be driven out of equilibrium in order for chirality to be maintained. We thus assume that the system is thermodynamically open, due to fluxes of matter in and out of the system.  

Let us also suppose that species entering the autocatalytic system are achiral but of high free energy, while the achiral species produced by the reactions involving the two D- and L-enantiomers have a lower free energy. 
In this regard,  the achiral species $\{A_a\}_{a=1}^{\NA}$ are of high free energy, and the achiral species $\{\tilde A_a\}_{a=1}^{\NAi}$ of low free energy.  The reaction networks are given by the following reactions:
\bea
&&{\rm A}_a +{\rm E}_i  \rightleftharpoons {\rm E}_j + {\rm E}_k \, , \label{react1} \\
&&{\rm A}_a +\bar{\rm E}_i  \rightleftharpoons \bar{\rm E}_j + \bar{\rm E}_k \, , \label{react2} \\
&&{\rm E}_i + \bar{\rm E}_j   \rightleftharpoons \tilde{\rm A}_b + \tilde{\rm A}_c  \, , \label{react3} 
\eea
where the enantiomer species are either ${\rm E}_m={\rm D}_m$ and $\bar{\rm E}_m={\rm L}_m$, or ${\rm E}_m={\rm L}_m$ and $\bar{\rm E}_m={\rm D}_m$ for 
each enantiomeric pair $m=i,j,k=1,2,\dots,\NC$; $a=1,2,\dots,\NA$; and $b,c=1,2,\dots,\NAi$.  Equations~(\ref{react1})-(\ref{react2})-(\ref{react3}) define a total of $2^{N_{\rm C}-1}$ inequivalent reaction networks differing by the permutations of D- and L-enantiomers for 
some enantiomeric pairs.  For given reaction rates, all these networks manifest similar dynamical behaviors.  Among them, the network with ${\rm E}_m={\rm D}_m$ and $\bar{\rm E}_m={\rm L}_m$ for all the pairs $m=1,2,\dots,N_{\rm C}$ is the direct generalization of Frank's model, considered below.

For our numerical implementation of this model,   
we focus on the fully irreversible regime, in which reactions (\ref{react1}), (\ref{react2}), and (\ref{react3}) only proceed forward due to the supply of achiral species with high free energy at the same concentration $A_0$. 
Thus, there are two main control parameters in this model: the supply concentration $A_0$ and the residence time $\tau$.

For one particular realization of these rate constants,  
Fig.~\ref{fig-frank}a shows the evolution of the concentrations of the species present in the system as function of time above the threshold concentration $A_0$, while Fig.~\ref{fig-frank}b shows the case below threshold.  In Fig.~\ref{fig-frank}a, we see that on long times the system converges towards a steady state, which is homogeneous and chiral. Only two species have been shown in these figures for clarity but the time evolution of their concentrations is typical of the evolution of all the other species: on long times, only one enantiomer is present, which is of the same chirality for all the species, while the other enantiomer reaches a vanishing concentration. Instead, in Fig.~\ref{fig-frank}b, all the species converge on long times towards a vanishing concentration.

\begin{figure}[h!]
\centering
\includegraphics[width=1\linewidth]{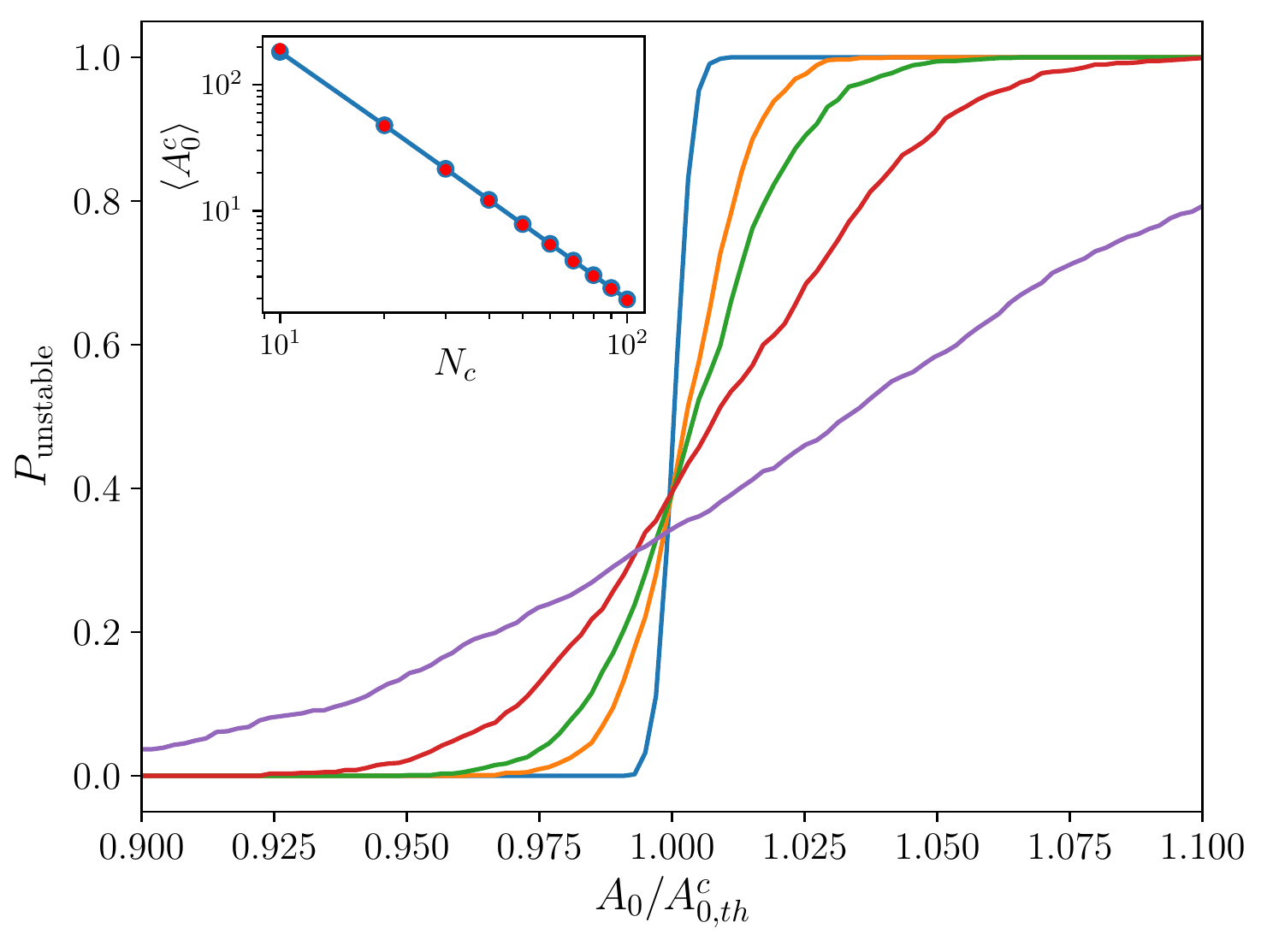}
\caption{Probability of instability of the racemic state for the generalized Frank model (\ref{react-ADDD-num})-(\ref{react-ALLL-num})-(\ref{react-AADL-num}). Probability of the initial racemic state to be unstable by mechanism (i) as function of the normalized value of the control parameter $A_0$ for the expanded Frank's model in the irreversible regime and for different values of number of chiral species~$N_{\rm C}$: $N_{\rm C}=10$ (magenta), $N_{\rm C}=20$ (red), $N_{\rm C}=30$ (green), $N_{\rm C}=40$ (yellow) and $N_{\rm C}=100$ (blue). The control parameter $A_0$ has been normalized by the theoretical threshold at the transition, defined by the equality in the inequality of Eq.~(\ref{criterion}). An average over a thousand realizations of the rate constants following a log-normal distribution such that $\langle k_+ \rangle = \langle \tilde k_- \rangle = 10^{-4}$ and $\sigma_{k_+} = \sigma_{\tilde k_-} = 2\times 10^{-4}$ has been performed. Inset: comparison between the observed control parameter value $A_0$ at the transition (red circles) with the theoretical prediction given by Eq.~(\ref{criterion}) (blue solid line) after averaging over $100$ realizations of the rate constants.}
\label{fig-instability}
\end{figure}

The case where all the rate constants would be identical can be treated analytically as done in (SI Appendix, section \ref{Instability_gen_Franck}), so
let us now instead assume that the rate constants $k_+$ of reactions (\ref{react1}) and (\ref{react2}) are taken according to a log-normal distribution \cite{davidi_birds-eye_2018}. 
We find that the spontaneous chiral symmetry breaking happens if the following criterion is satisfied,
\be\label{criterion}
\langle k_+ \rangle \, \tau \, \NA\, A_0 > \frac{2}{\NC(\NC+1)} \, ,
\ee
{\it i.e.}, the residence time multiplied by the total concentration $\NA\, A_0$ of achiral species supplied to the system must exceed a threshold determined by the average rate constant $\langle k_+ \rangle$ of autocatalysis and the number $\NC$ of chiral species in the reaction network. 
We have tested this result by performing a linear stability analysis of the racemic steady state.  Simulations show that instability is due to an isolated top eigenvalue, and confirm the criterion (\ref{criterion}) when the distribution of the rate constants is not too broad as shown in the inset of Fig.~\ref{fig-instability}. For a very broad distribution, the threshold is pushed towards higher value than predicted by Eq.~(\ref{criterion}), but the transition still occurs at sufficiently large $\NC$ (SI Appendix, Fig.~\ref{verif_criterion}). Importantly, 
the transition becomes sharp as the number of chiral species increases as shown in Fig.~\ref{fig-instability}.
Thus, the random matrix theory argument holds and the mechanism (i) is confirmed, although 
the eigenvalues of the Jacobian matrix 
 do not cover uniformly a circle (SI Appendix, Fig.~\ref{fig-vp_jacob_zoom}) due to the difference of statistics between the diagonal and the off-diagonal elements (SI Appendix, section \ref{Instability_gen_Franck}). In this case, the mechanism (ii) is found not to be relevant.
Furthermore, the multiplication of chiral species is also multiplying the number of reaction networks manifesting similar chiral symmetry breaking, but with either the D- or the L-enantiomer for the different chiral species.

\subsection*{Discussion} Let us now come back to our evaluation of the cross-over to chirality in the context of the origin of life. First of all, we observe that among the 20 natural amino acids, the only one that is achiral$-${\it i.e.}, glycine$-$is also the smallest containing only 5 heavy atoms. Secondly, the first chiral molecule observed in space is propylene oxide, containing 4 heavy atoms \cite{mcguire_discovery_2016}. Thirdly, among the 11 carboxylic acids of the Krebs cycle, the majority of them, 9 lie in the range between 8 and 13 heavy atoms, only two of them are smaller, acetate and pyruvate.
In fact, the Krebs cycle appears to function precisely at the border between the world of achiral and small molecules and that of large and chiral molecules \cite{smith_origin_2016}. 
The emergence of the Krebs cycle thus represents a major step, which facilitates the synthesis of a large number of long chiral molecules \cite{morowitz_origin_2000}. 

We presented a scenario that explains why a large complex molecular system tends to become chiral. Based on fundamental properties of phase transitions, confirmed by numerical simulations of a non-equilibrium reaction network, our scenario is both general and robust. Details of the reaction network should not matter, nor the precise way in which the system is driven out of equilibrium, provided the system is large enough.  In addition, the homochiral state of our model does not need to be 
all D or L across all species, in agreement with the observation that 
for instance, amino-acids are L-chiral, but sugars are D-chiral.
Moreover, our reaction scheme needs not satisfy mass-action law, and there is also no  requirement that the system be fully well-mixed, it could be compartmentalized. As an illustration of this idea, we study two diffusively coupled chemical reactors, identical to the one considered so far (SI Appendix, section \ref{TDCC}).
For low coupling, the two compartments undergo separately the homochiral transition, while at high coupling, they reach the same homochiral state \cite{jafarpour_noise-induced_2017}. Compartmentalized systems enrich the scenarios for the transition to homochirality, because on one hand, as shown recently, compartmentalization significantly broadens the diversity of available autocatatalytic networks which can be built with a limited number of compounds \cite{blokhuis_universal_2020}, and on the other hand, in such systems, the number of species is effectively increased, which favors our mechanism.

Our scenario thus offers a universal pathway towards homochirality potentially unifying many previous approaches on this issue. In the context of the origin of life, we also find that there is no need for very long and complex molecules for this homochiral state to emerge. The transition can already occur in a prebiotic world containing molecules with about $10$ heavy atoms.

\subsection*{Materials and Methods}
\paragraph{Chemoinformatics of chirality} A study of the scaling of the number of chiral and achiral species with the number of atoms for monosubstituted alkanes and alkanes stereoisomers is provided in SI Appendix section \ref{Main}, together with details on the chemoinformatic analysis of the PubChem database \cite{pubchem_database}.

\paragraph{Chiral symmetry breaking for a general reaction scheme} The spontaneous symmetry breaking of chirality can be described in the framework of kinetics.
The reaction network is supposed to involve achiral and chiral species at the concentrations
\be
{\bf c}=
\left( \begin{array}{c}
{\bf c}_{\rm D} \\
{\bf c}_{\rm A} \\
{\bf c}_{\rm L}
\end{array}
\right) ,
\ee
which respectively denote the concentrations of D-enantiomers, achiral species, and L-enantiomers.  If $N_{\rm D}$, $N_{\rm A}$, and $N_{\rm L}$ denote the respective numbers of these species, the system is described in terms of $N_{\rm S}=N_{\rm D}+N_{\rm A}+N_{\rm L}=N_{\rm A}+2N_{\rm C}$ concentrations with $N_{\rm C}\equiv N_{\rm D}=N_{\rm L}$.

In an open system, the time evolution of these concentrations is ruled by the $N_{\rm S}$ kinetic equations in Eq.~(\ref{kin_eqs}) with ${\bf F}({\bf c})=\pmb{\nu}\cdot{\bf w}({\bf c})$,
expressed in terms of the matrix $\pmb{\nu}$ of stoichiometric coefficients, the set ${\bf w}({\bf c})$ of net reaction rates $w_r({\bf c})=w_{+r}({\bf c})-w_{-r}({\bf c})$ with $r=1,2,...,N_{\rm R}$, the supply concentrations ${\bf c}_0$, and the residence time $\tau$ of the species in the system. The system is closed if the residence time is infinite, in which case the last term drops in 
Eq.~(\ref{kin_eqs}.

The mirror symmetry of the system corresponds to the following exchange of concentrations of D- and L-enantiomers,
\be
{\boldsymbol{\mathsf S}}_{\bf c}\cdot{\bf c}=
\left( \begin{array}{ccc}
0 & 0 & {\boldsymbol{\mathsf I}} \\
0 & {\boldsymbol{\mathsf I}} & 0\\
{\boldsymbol{\mathsf I}}  & 0 & 0
\end{array}
\right) \cdot
\left( \begin{array}{c}
{\bf c}_{\rm D} \\
{\bf c}_{\rm A} \\
{\bf c}_{\rm L}
\end{array}
\right) 
=
\left( \begin{array}{c}
{\bf c}_{\rm L} \\
{\bf c}_{\rm A} \\
{\bf c}_{\rm D}
\end{array}
\right) ,
\ee
written in terms of the $N_{\rm S}\times N_{\rm S}$ matrix such that ${\boldsymbol{\mathsf S}}_{\bf c}^2 = {\boldsymbol{\mathsf I}}$, where ${\boldsymbol{\mathsf I}}$ denotes the corresponding identity matrix. Since the rate constants take equal values for mirror-symmetric reactions, the reaction rates have the symmetry
\be
{\bf w}({\boldsymbol{\mathsf S}}_{\bf c}\cdot{\bf c}) = {\boldsymbol{\mathsf S}}_{\bf w}\cdot{\bf w}({\bf c})
\ee
with some $N_{\rm R}\times N_{\rm R}$ matrix satisfying ${\boldsymbol{\mathsf S}}_{\bf w}^2={\boldsymbol{\mathsf I}}$.  As a consequence of the mirror symmetry, the  $N_{\rm S}\times N_{\rm R}$ matrix of stoichiometric coefficients obeys the following symmetry relation
\be
{\boldsymbol{\mathsf S}}_{\bf c}\cdot\pmb{\nu}\cdot{\boldsymbol{\mathsf S}}_{\bf w}=\pmb{\nu} \, .
\ee
We note that the kinetic equations may also be written in the following form,
\bea
&& \frac{d{\bf c}_{\rm D}}{dt} = {\bf F}_{\rm D}({\bf c}_{\rm D},{\bf c}_{\rm A},{\bf c}_{\rm L}) + \frac{1}{\tau}({\bf c}_{{\rm D}0}-{\bf c}_{\rm D}) \, , \\
&& \frac{d{\bf c}_{\rm A}}{dt} = {\bf F}_{\rm A}({\bf c}_{\rm D},{\bf c}_{\rm A},{\bf c}_{\rm L}) + \frac{1}{\tau}({\bf c}_{{\rm A}0}-{\bf c}_{\rm A}) \, , \\
&& \frac{d{\bf c}_{\rm L}}{dt} = {\bf F}_{\rm L}({\bf c}_{\rm D},{\bf c}_{\rm A},{\bf c}_{\rm L}) + \frac{1}{\tau}({\bf c}_{{\rm L}0}-{\bf c}_{\rm L}) \, , 
\eea
where the mirror symmetry is expressed as
\be\label{mirror_symm}
{\bf F}_{\rm D}({\bf x},{\bf y},{\bf z}) = {\bf F}_{\rm L}({\bf z},{\bf y},{\bf x}) \qquad\mbox{and}\qquad
{\bf F}_{\rm A}({\bf x},{\bf y},{\bf z}) = {\bf F}_{\rm A}({\bf z},{\bf y},{\bf x}) \, .
\ee

The symmetry can be explicitly broken by input concentrations such that ${\boldsymbol{\mathsf S}}_{\bf c}\cdot {\bf c}_0 \ne {\bf c}_0$.  However, the equations remain symmetric if the condition ${\boldsymbol{\mathsf S}}_{\bf c}\cdot {\bf c}_0 = {\bf c}_0$ holds.  In this case, the racemic mixture characterized by equal concentrations of D- and L-enantiomers,
\be\label{racemic-subspace}
{\bf c}_{\rm D}={\bf c}_{\rm L} \, , 
\ee
is maintained during the time evolution of the reaction network, if the dynamics
\bea
&& \frac{d{\bf c}_{\rm A}}{dt} = {\bf F}_{\rm A}({\bf c}_{\rm D},{\bf c}_{\rm A},{\bf c}_{\rm D}) + \frac{1}{\tau}({\bf c}_{{\rm A}0}-{\bf c}_{\rm A}) \, , \\
&& \frac{d{\bf c}_{\rm D}}{dt} = {\bf F}_{\rm D}({\bf c}_{\rm D},{\bf c}_{\rm A},{\bf c}_{\rm D}) + \frac{1}{\tau}({\bf c}_{{\rm D}0}-{\bf c}_{\rm D})\, , 
\eea
is stable in the racemic subspace~(\ref{racemic-subspace}).  In order to investigate this issue, we introduce the variables
\be
{\bf x} \equiv \frac{1}{2} ({\bf c}_{\rm L}-{\bf c}_{\rm D})
\ee
characterizing deviations with respect to the racemic subspace and we perform the linear stability analysis for infinitesimal deviations $\delta{\bf x}$ with respect to the racemic subspace.  These deviations are ruled by the following set of linear equations:
\be
\frac{d}{dt} \delta{\bf x} = \left({\boldsymbol{\mathsf J}}_{\rm DD}-{\boldsymbol{\mathsf J}}_{\rm DL}-\frac{1}{\tau}\right)\cdot\delta{\bf x} + \frac{1}{\tau}\, \delta{\bf x}_0 \, ,
\ee
where
\be
{\boldsymbol{\mathsf J}}_{\rm DD} \equiv \frac{\partial{\bf F}_{\rm D}}{\partial{\bf c}_{\rm D}} =\frac{\partial{\bf F}_{\rm L}}{\partial{\bf c}_{\rm L}}\qquad\mbox{and}\qquad  {\boldsymbol{\mathsf J}}_{\rm DL} \equiv \frac{\partial{\bf F}_{\rm D}}{\partial{\bf c}_{\rm L}} =\frac{\partial{\bf F}_{\rm L}}{\partial{\bf c}_{\rm D}}\, .
\ee
Note that the chiral symmetry conditions, namely Eq.~(\ref{mirror_symm}) have been used to derive these equations.
We suppose that $\delta{\bf x}_0=0$, so that there is no explicit symmetry breaking.  
Moreover,  the dynamics in the racemic subspace is assumed to have a steady state. We use the notations 
${\boldsymbol{\mathsf J}}= {\boldsymbol{\mathsf J}}_{\rm DD}-{\boldsymbol{\mathsf J}}_{\rm DL}$ and 
 ${\boldsymbol{\mathsf M}}\equiv  {\boldsymbol{\mathsf J}}-{\boldsymbol{\mathsf I}}/\tau$.
The $N_{\rm C}\times N_{\rm C}$ ${\boldsymbol{\mathsf M}}$ matrix controls the linear stability of the racemic steady state.  It is asymptotically stable if all the eigenvalues $\{\lambda_i\}$ of that matrix have a negative real part: ${\Re}\, \lambda_i<0$ for all $i=1,2,...,N_{\rm C}$.  The racemic mixture is unstable if at least one of its eigenvalues has a positive real part.


\paragraph{Generalized Frank's model} For numerical investigations, we consider the model (\ref{react1})-(\ref{react2})-(\ref{react3}) with ${\rm E}_i={\rm D}_i$ and $\bar{\rm E}_i={\rm L}_i$ for all $i=1,2,\dots,N_{\rm C}$ in the irreversible regime with $\NA = \NAi = 1$ and $N\equiv \NC \gg 1$. Moreover, we suppose that the initial concentrations $D_{m0}=L_{m0}=\tilde A_0=0$ for all species $m$.  Since $D_{m0}=L_{m0}$, there is no explicit chiral symmetry breaking caused by non-racemic inflow from the environment.  In this regime, the reaction network reads
\bea
&&{\rm A} +{\rm D}_i  \rightarrow {\rm D}_j + {\rm D}_k \, , \label{react-ADDD-num} \\
&&{\rm A} +{\rm L}_i \rightarrow {\rm L}_j + {\rm L}_k \, , \label{react-ALLL-num} \\
&&{\rm D}_i + {\rm L}_j   \rightarrow 2\, \tilde{\rm A}  \, , \label{react-AADL-num} 
\eea
where $i,j,k=1,2,\dots,N_{\rm C}$.  

Now, the net reaction rates are given by
\bea
&& w_{ijk}^{\rm (D)} = k_{+ijk} \, A \, D_i  \qquad\mbox{with}\qquad j\le k\, , \label{rate-ADDD-num}\\
&& w_{ijk}^{\rm (L)} = k_{+ijk} \, A \, L_i  \qquad\mbox{with}\qquad j\le k \, , \label{rate-ALLL-num} \\
&& \tilde w_{ij} = \tilde k_{-ij} \, D_i \, L_j \, ,\label{rate-AADL-num}
\eea
where $\tilde k_{-ij}=\tilde k_{-ji}$ because of the mirror symmetry. 
Rate constants are randomly 
distributed according to a log-normal distribution as explained in SI Appendix, section \ref{GFMSM}.

The kinetic equations have thus the following form,
\bea
\dot{A} &=& - \sum_{ijk\atop j\le k} w_{ijk}^{\rm (D)} - \sum_{ijk\atop j\le k} w_{ijk}^{\rm (L)} + \frac{1}{\tau} (A_{0}-A) \, , \label{eq-A-num} \\
\dot{D}_m &=& \sum_{ijk\atop j\le k} \nu_{m,ijk}\, w_{ijk}^{\rm (D)}  -\sum_{i} \tilde w_{mi} -\frac{1}{\tau} D_m \, , \label{eq-Dm-num} \\
\dot{L}_m &=&\sum_{ijk\atop j\le k} \nu_{m,ijk}\, w_{ijk}^{\rm (L)} -\sum_{i} \tilde w_{im} -  \frac{1}{\tau} L_m \, , \label{eq-Lm-num} \\
\dot{\tilde A} &=& 2\, \sum_{ij}  \tilde w_{ij} - \frac{1}{\tau} \tilde A \, , \label{eq-Ai-num} 
\eea
with $\nu_{m,ijk} \equiv -\delta_{mi} + \delta_{mj} + \delta_{mk}.$


\paragraph{Numerical simulations of the reaction network} The above equations for the fully irreversible model, 
have been simulated using a Runge-Kutta algorithm of second order. The numerical integration of the kinetic equations has been performed by setting $\tau=1$, meaning that we take $\tau$ as time unit.
At the initial time, we assume that there is a very small imbalance between the two enantiomers of 
given species, characterized by the small parameter $\epsilon$, which is homogeneous among all the species.

The integration of the ordinary differential equations allows us to determine the threshold of instability, as well as every asymptotically stable solution$-$in particular, the racemic solution with $D_i=L_i$ below the threshold of instability.  This threshold can be determined by increasing the control parameter $A_0$ until the solution of the equations is no longer racemic, giving the critical value of the threshold $A_{0{\rm c}}$ for the transition-breaking chiral symmetry inside the system.

With this dynamics, we observe that the system never converges towards a non-trivial racemic state where the concentrations of the two enantiomers of a given species would be non-zero and equal to each other. 
Thus, starting with a state with a small enantiomeric excess, 
we either reach the trivial racemic state or a homochiral state. 
For this reason, Fig.~\ref{fig-instability} has been made by studying the stability of the trivial racemic state using many random realizations of the rate constants $k_{+ijk}$ following a log-normal distribution, which is more efficient numerically than a time integration of the equations of motion.
{\subsection*{Further materials} 
In SI Appendix, section \ref{Jacobian-matrix}, we study the properties of the Jacobian matrix, and deduce from them in section \ref{GIC} a general instability criterion of the racemic state based on random matrix theory.
In section \ref{GFMSM}, we present the reversible generalized Frank model and we analyze its properties for uniform rate constants. In section \ref{Instability_gen_Franck}, we derive the instability criterion for the generalized Frank model with random rate constants. In section \ref{TDCC}, we study two diffusively coupled compartments.}
%
%
%

\subsection*{Acknowledgments}
The authors acknowledge fruitful discussions with A. Blokhuis and A. Duprat. L. Leibler and Y. Geerts are thanked for their helpful comments. PG acknowledges the financial support of the Universit\'e Libre de Bruxelles (ULB) and the Fonds de la Recherche Scientifique-FNRS under the grant PDR T.0094.16 for the project ``SYMSTATPHYS''. DL acknowledges support from Agence Nationale de la Recherche (ANR-10-IDEX-0001-02, IRIS OCAV) and (ANR-11-LABX-0038, ANR-10-IDEX-
0001-02).


\bibliography{Chiral}
\bibliographystyle{ieeetr}


\makeatletter 
\renewcommand{\thefigure}{S\@arabic\c@figure} 
\setcounter{figure}{0}

\makeatletter 
\renewcommand{\theequation}{S\@arabic\c@equation} 
\setcounter{equation}{0}

\onecolumn
\newpage
\renewcommand\appendixpagename{\centering Supplementary Materials}
\appendixpage
\vspace{1cm}

\appendix

\makeatletter 
\renewcommand{\thesection}{S\@arabic\c@section} 
\setcounter{section}{0}

\section{The multiplication of chiral molecules with their number of atoms}
\label{Main}

The number of species that are chiral are observed to increase faster with their number of atoms, than the number of achiral species, in spite of the fact that diatomic and triatomic molecules are achiral and only tetratomic molecules can be chiral (in their ground electronic state).  In particular, studies of alkane and monosubstituted alkane stereoisomers \cite{F07} show that the numbers of chiral $C_k$ and achiral $A_k$ species are growing exponentially with their number $k$ of carbon atoms according to
\be
C_k \sim \Lambda^k \qquad\mbox{and}\qquad A_k \sim \Lambda^{k/2} \qquad\mbox{with} \qquad \Lambda \simeq 3.287112
\ee
for $k\gg 1$. As a consequence, chiral molecules become overwhelmingly dominant for a large enough number of atoms.

Now, the question is to determine when the crossover occurs between a world of small molecules dominated by achiral species and a world dominated by chiral molecules, which is likely to become homochiral by spontaneous symmetry breaking induced in the non-equilibrium reaction network.

In order to answer this question, we have investigated the fractions of achiral and chiral species as functions of the number of atoms in each molecule.  Such fractions can be defined by counting the pairs of enantiomers either once or twice.  The numbers of achiral and chiral species with $n$ heavy atoms being respectively denoted $A_n$ and $C_n$, on the one hand, the fractions defined by counting once the pairs of enantiomers are given by
\be
f_n^{\rm (A)} \equiv \frac{A_n}{A_n+C_n} \qquad\mbox{and}\qquad f_n^{\rm (C)} \equiv \frac{C_n}{A_n+C_n} ,
\ee
such that $f_n^{\rm (A)}+f_n^{\rm (C)}=1$, in which case, the crossover happens for $n_1$ atoms in the molecule such that
\be
A_{n_1}= C_{n_1} \, .
\ee
On the other hand, the fractions defined by counting twice the pairs of enantiomers are given by
\be
\tilde f_n^{\rm (A)} \equiv \frac{A_n}{A_n+2C_n} \qquad\mbox{and}\qquad \tilde f_n^{\rm (C)} \equiv \frac{2C_n}{A_n+2C_n} ,
\ee
such that $\tilde f_n^{\rm (A)}+\tilde f_n^{\rm (C)}=1$ in which other case, the crossover happens for $n_2$ atoms in the molecule such that
\be
A_{n_2}=2 C_{n_2} \, .
\ee
Since $C_n$ becomes larger than $A_n$ as $n$ increases, we should expect that $n_2 < n_1$.

\subsection{Monosubstituted alkane stereoisomers}

\begin{figure}[h]
\centerline{\scalebox{0.4}{\includegraphics{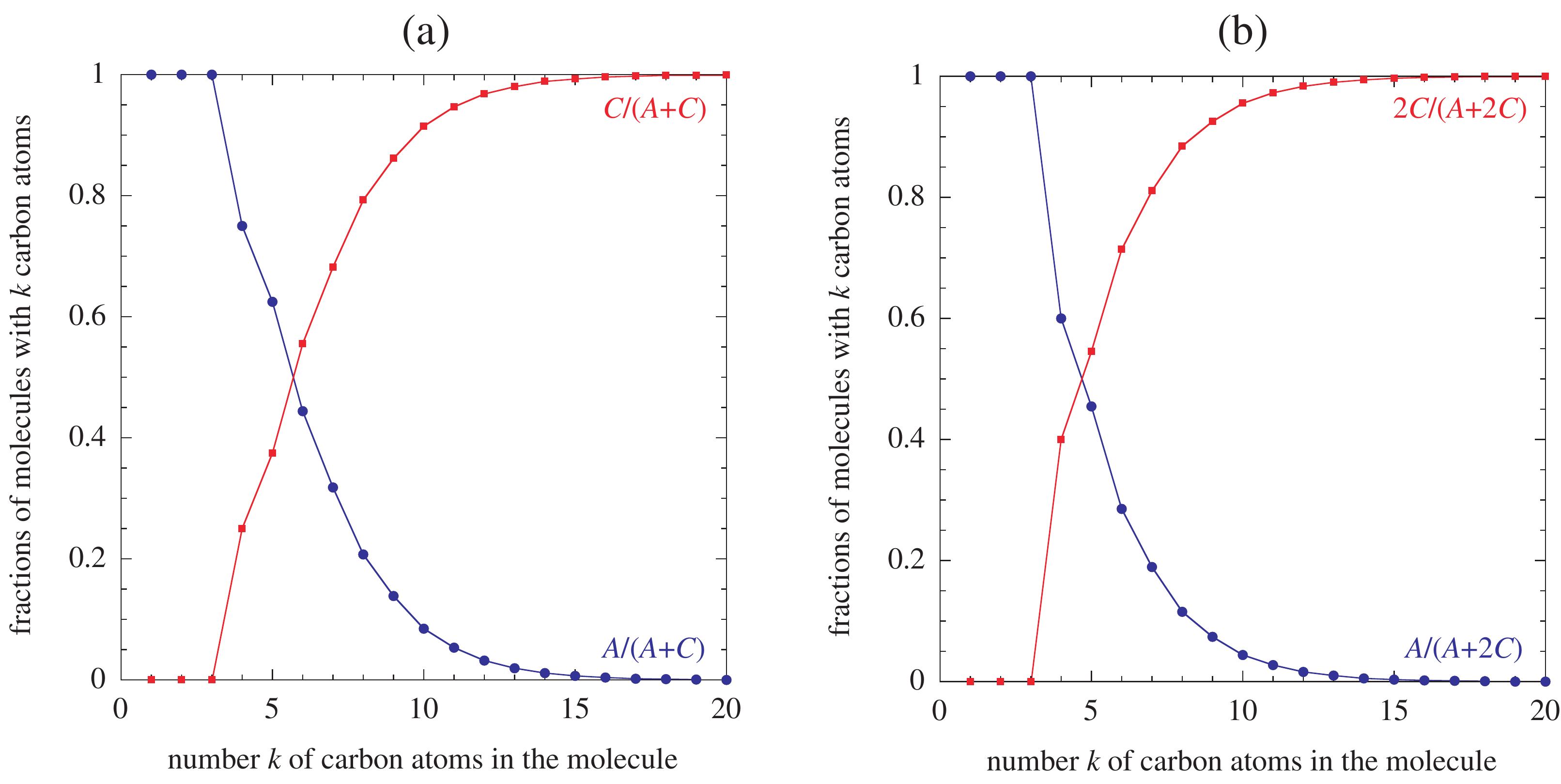}}}
\caption{Fractions of achiral and chiral stereoisomers of monosubstituted alkanes versus the number $k$ of carbon atoms in the molecule, counting (a) once and (b) twice the pairs of enantiomers.  The data are from Ref.~\cite{F07}.}
\label{fig2}
\end{figure}

Achiral and chiral monosubstituted alkanes ${\rm C}_k{\rm H}_{2k+1}{\rm X}$ have been enumerated in Ref.~\cite{F07}.  In particular, Table 1 of Ref.~\cite{F07} gives the numbers of achiral $A_k$ and chiral $C_k$ monosubstituted alkanes as stereoisomers versus the number $k$ of carbon atoms they contain and this up to $k=100$.  Using these data, the fractions of achiral and chiral stereoisomers have been obtained and they are shown in Fig.~\ref{fig2} by counting once or twice the enantiomeric pairs.


We observe that the crossover occurs at $k_1\simeq 5.7$ or $k_2\simeq 4.7$, depending on whether the pairs of enantiomers are counted once or twice.  As expected, we have that $k_2<k_1$.  Now, the result is that the crossover happens for a relatively small number of carbon atoms.  Here, the carbon atoms and the substituted atom X determine the geometry of the molecule.  Here, the temperature is supposed to be high enough such that hydrogen atoms rotate and vibrate fast enough that the chirality is determined by the skeleton of the carbon and  X atoms.

\subsection{Alkane stereoisomers}

\begin{figure}[h]
\centerline{\scalebox{0.4}{\includegraphics{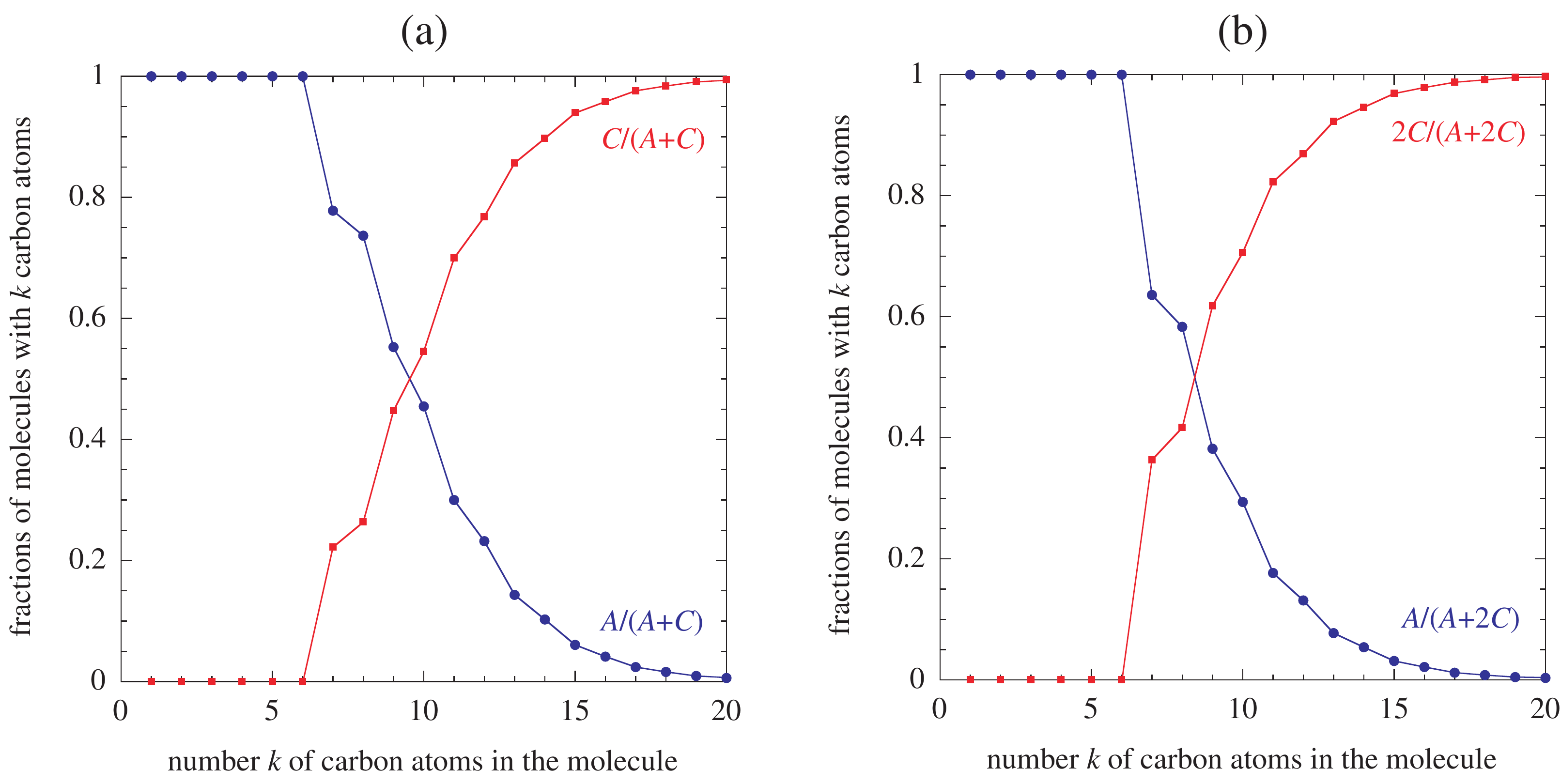}}}
\caption{Fractions of achiral and chiral stereoisomers of alkanes versus the number $k$ of carbon atoms in the molecule, counting (a) once and (b) twice the pairs of enantiomers.  The data are from Ref.~\cite{fujita_alkanes_2007}.}
\label{fig3}
\end{figure}

The stereoisomers of alkanes ${\rm C}_k{\rm H}_{2k+2}$ have also been studied and the enumeration of achiral and chiral alkanes is given in Ref.~\cite{fujita_alkanes_2007}.  Table 3 of Ref.~\cite{fujita_alkanes_2007} gives the numbers of achiral $A_k$ and chiral $C_k$ alkanes as stereoisomers versus the number $k$ of carbon atoms they contain and this up to $k=100$.  The fractions of achiral and chiral stereosiomers obtained with these data are plotted in Fig.~\ref{fig3}.


Here, we see that the crossover occurs at $k_1\simeq 9.5$ if the pairs of enantiomers are counted once and at $k_2\simeq 8.4$ if they are counted twice. Again, the crossover happens for a relatively small number of carbon atoms (which are the atoms determining the molecular geometry).  The crossover happens for somewhat larger molecules because alkanes have molecular structures that are more symmetric than in the presence of one substitution, thus delaying the crossover as the number of determining atoms increases.

\subsection{Chemical Universe}

In Ref.~\cite{FR07}, all the possible molecules up to 11 atoms of C, N, O, and F were generated by considering simple valency, chemical stability, and synthetic feasibility rules, and they were collected in a database  containing 26.4 million molecules and 110.9 million stereoisomers.  Fig.~5 of Ref.~\cite{FR07} shows the fractions of achiral and chiral molecules in the database as a function of their size characterized by the number of heavy atoms.  Here, the crossover happens for $n_1\simeq 8.5\pm 0.1$.  This virtual exploration of the chemical universe clearly demonstrates the prevalence of chirality for large enough molecules.


\subsection{Analysis of the PubChem database}

\subsubsection{Raw data}

\begin{figure}[h!]
\centering
\begin{subfigure}[t]{0.4\textwidth}
\centering
\includegraphics[scale=0.53]{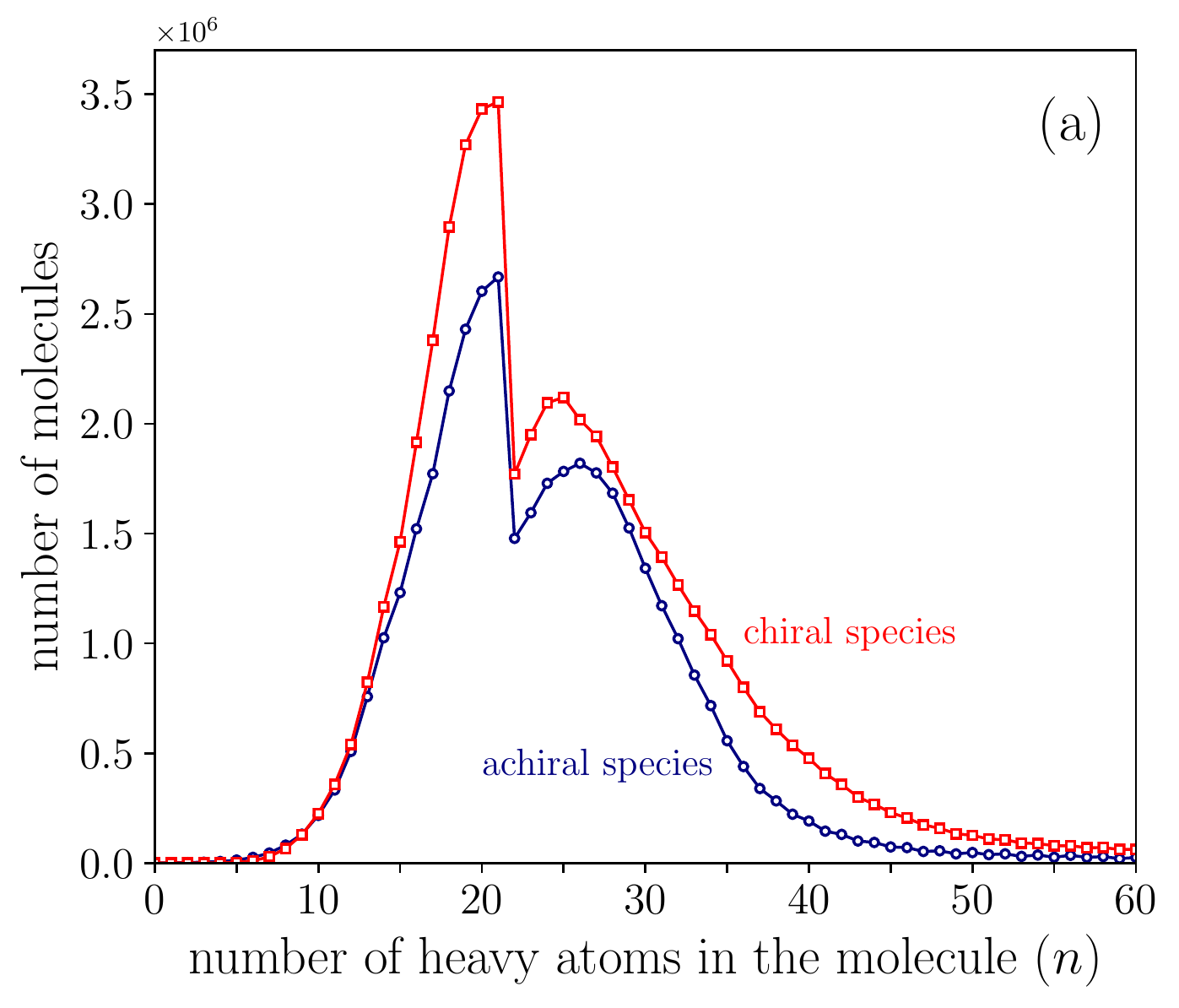}
\caption{}
\end{subfigure}%
~
\begin{subfigure}[t]{0.4\textwidth}
\centering
\includegraphics[scale=0.53]{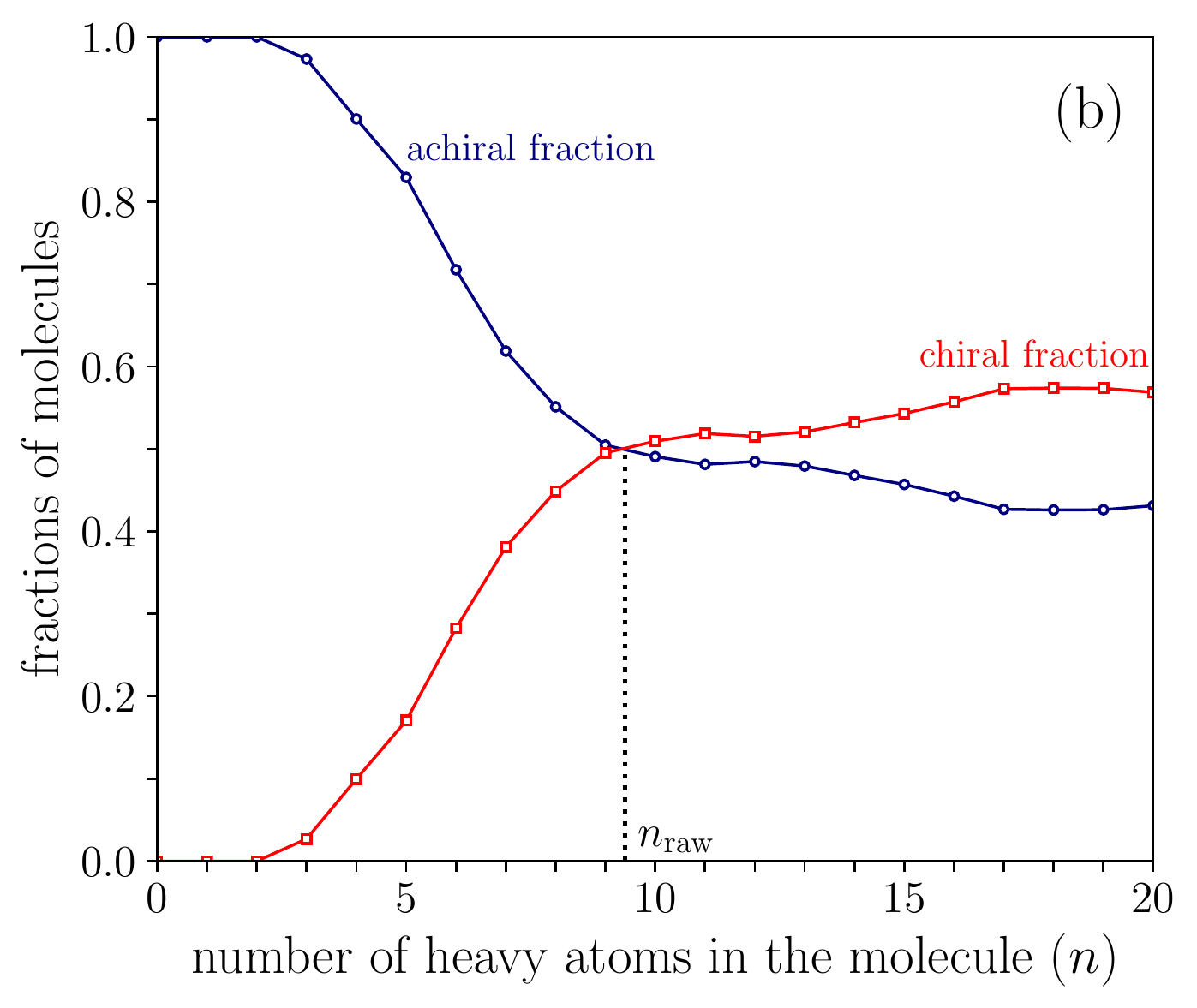}
\caption{}
\end{subfigure}
\caption{(a) Total number of achiral and chiral species in the raw PubChem database containing about $139$ millions of species. (b) Fractions of chiral and achiral molecules containing $n \leq 20$ heavy atoms. In this case, only $N = 33,563,343$ molecules with $n \leq 20$ heavy atoms were analyzed after the specific selection. The figure on the right shows an intersection at $n_{\rm raw} \simeq 9.4$.}
\label{fig:raw_pubchem}
\end{figure}

The raw database of PubChem contains $139$ millions of species. In the following, we will restrict our analysis to species which contain less than $20$ heavy atoms ({\it i.e.}, atoms heavier than hydrogen). There are two reasons for this choice, on one hand the statistics becomes more limited for molecules much longer than $20$, and on the other hand, there is a discontinuity in the number of achiral and chiral molecules in the PubChem database
as shown in Fig.~\ref{fig:raw_pubchem}a. We have contacted the curators of the database, but there is no information currently available about the origin of this discontinuity.  In any case, we should avoid this problem by staying below $20$.


From the complete dataset downloaded from PubChem, $91,606,016$ molecules were analyzed after rejection of compounds with isotopic elements, multiple components or incomplete data on bond structure. 
From the $33,563,343$ molecules with less than 21 heavy atoms, $18,705,878$ molecules (55.7 \%) are chiral, and $1,376,672$ chiral molecules have no stereocenters (7.4 \% of the chiral molecules with less than 21 heavy atoms) thus for these molecules their chirality depends only on their non planar geometry. 

As shown in Fig.~\ref{fig:raw_pubchem}b, the analysis of the $33,563,343$ molecules with less than 21 heavy atoms of the database in terms of their fraction of chiral and non-chiral species 
shows a crossover around $n_{\rm raw} \simeq 9.4$.
A crossover in this region is coherent with an increase in the number of stereoisomers per molecule for molecules of this length.

\subsubsection{Methods of generation of stereoisomers and enantiomers}

Chiral species in the PubChem database were detected using the chiral flag present in the list of SDF files which contains information about the structure of the molecules in the database. 

For the  generation of enantiomers, a list of non-canonical SMILES formulas was built, which contains information about defined stereocenters. For each chiral molecule with an available SMILES formula ({\it e.g.}, C[\textbf{C@@}](CC1CC=C(C(=C1)O)O)(C(=O)O)N), one generates a mirror image of it ({\it i.e.}, C[\textbf{C@}](CC1CC=C(C(=C1)O)O)(C(=O)O)N for the latter example), and then one searches it in the list. If it is not found, then it is added to the list in the expanded database. However, for chiral centers that do not explicitly appears in the SMILES formula, one cannot find them and reverse them with a simple method and this creates an uncertainty in the final number of chiral molecules due to missed generated enantiomers. Thus a fraction of enantiomers cannot be generated due to incomplete data in the PubChem database.

For the generation of stereoisomers, one looks whether the database contains the theoretical maximum number of stereoisomers for a given species, which can be evaluated from the number of defined stereocenters. If all the stereoisomers are not present, which is frequent for molecules containing several stereocenters, the database is expanded so that each chiral species has the maximum possible number of stereoisomers ({\it i.e.}, $2^{n^*}$ with $n^*$ the number of stereocenters in the molecule). However, this procedure does not count properly the meso forms, which should be labeled as achiral although they contain stereocenters due to an internal symmetry.



\begin{figure}[h!]
\centering
\includegraphics[scale=0.5]{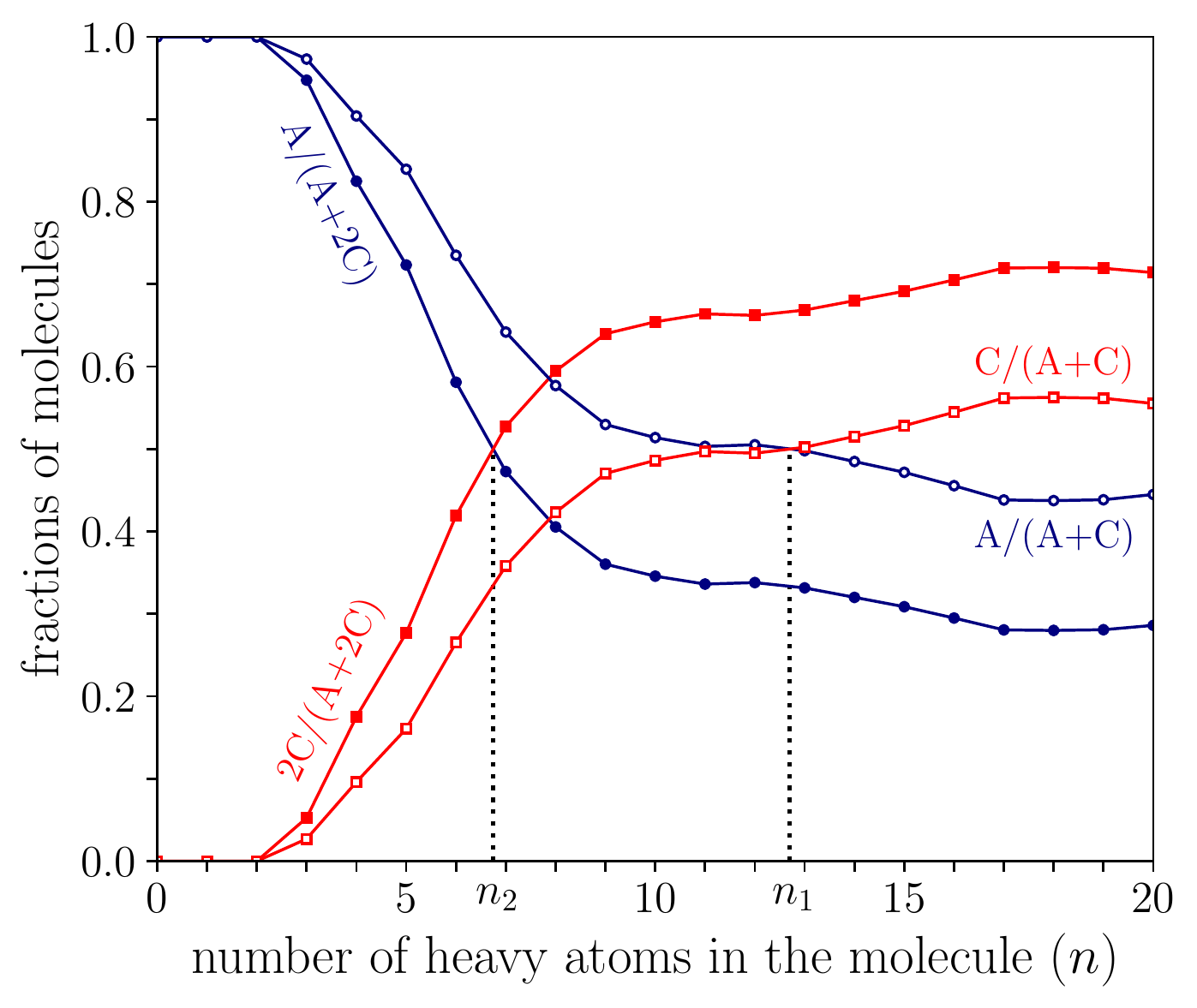}
\caption{Analysis of the database expanded in enantiomers, with $N = 50,252,957$ molecules in data ({\it i.e.}, $16,689,614$ enantiomers were generated) with $n \leq 20$ heavy atoms. The intersection occurs at $n_{2} \simeq 6.7$ for if both enantiomers are considered and $n_{1} \simeq 12.7$ if only one enantiomer is considered.}
\label{fig:enantiomer_pubchem}
\end{figure}

Now, after generating enantiomers as explained previously, one obtains Fig.~\ref{fig:enantiomer_pubchem}.


Then, stereoisomers were generated using the procedure described previously, giving Fig.~1a of the main article where 47,452,700 steoisomers have been added to complete the PubChem dataset. In this case, the intersection occurs at $n_{2} \simeq 6.4$ for if both enantiomers are considered and $n_{1} \simeq 8.1$ if only one enantiomer is considered.



\subsubsection{Error bars}

In this subsection, we explain how error bars were obtained in the graph of achiral and chiral fractions. For molecules with $i$ atoms, the number of chiral molecules is denoted ${\cal N}_i$ and the one of achiral molecules ${\cal M}_i$.  These numbers are taken as independent Poisson distributions with a parameter given by their mean number, {\it i.e}., by ${\cal N}_i$ and ${\cal M}_i$ themselves.  The fraction of chiral molecules is given by
\be
x_i = \frac{{\cal N}_i}{{\cal M}_i+{\cal N}_i} \, .
\ee
The error on $x_i$ thus reads
\be
\sigma_{x_i} = x_i (1-x_i) \left( \frac{1}{\sqrt{{\cal M}_i}}+\frac{1}{\sqrt{{\cal N}_i}}\right) .
\ee
The model does not capture systematic errors, but only the statistical errors in the counting of ${\cal N}_i$ or ${\cal M}_i$.

\section{Separating the mean from fluctuations in the Jacobian matrix}
\label{Jacobian-matrix}


We now study the properties of the $N\times N$ Jacobian matrix ${\boldsymbol{\mathsf J}}$ with $N=\NC$ introduced in the Materials and Methods of the main text. For large complex networks, this matrix may be supposed to be random because of fluctuations in the values of its elements for the different reactions and species.
Let us separate the mean of the elements of ${\boldsymbol{\mathsf J}}$ from their fluctuations in the following way:
\be\label{J-mu-G}
{\boldsymbol{\mathsf J}} = \mu \, {\bf 1} + \sigma\, {\boldsymbol{\mathsf G}} 
\ee
where $\bf 1$ is the matrix full of elements equal to $1$ and ${\boldsymbol{\mathsf G}}$ has elements distributed according to independent Gaussian distributions of zero mean and unit variance:
\be\label{G-dfn}
\langle G_{ab} \rangle = 0 \, , \qquad \langle G_{ab} G_{cd}\rangle = \delta_{ac}\delta_{bd} \, .
\ee
Accordingly, the mean value of Eq.~(\ref{J-mu-G}) gives
\be\label{mu-J-aver}
\langle{\boldsymbol{\mathsf J}}\rangle = \mu \, {\bf 1} \, ,
\ee
allowing us to determine the parameter $\mu$ as the mean value of the elements of the matrix $\boldsymbol{\mathsf J}$:
\be\label{mu}
\mu = \frac{1}{N^2} \sum_{a,b=1}^N J_{ab} \, .
\ee
Moreover, the parameter $\sigma$ can be evaluated by the root mean square of the matrix elements,
\be\label{sigma}
\sigma = \frac{1}{N} \left[\sum_{a,b=1}^N (J_{ab}-\mu)^2\right]^{1/2} ,
\ee
as a consequence of Eq.~(\ref{G-dfn}).

The matrix full of ones has the eigenvalues $\{N,0,0,...,0\}$ and it can be diagonalized with an orthogonal transformation ${\boldsymbol{\mathsf O}}$ composed of the eigenvectors ${\bf v}=\left\{\cos[2\pi k(m-1)/N]\right\}_{k=1}^N$ with $m=1,2,...,N$ (after their normalization).  The eigenvector corresponding to the eigenvalue equal to $N$ is thus given by  $m=1$ in the expression of the latter eigenvectors.  Accordingly,
\bea
{\boldsymbol{\mathsf J}}'={\boldsymbol{\mathsf O}}^{\rm T} \cdot {\boldsymbol{\mathsf J}} \cdot{\boldsymbol{\mathsf O}}&=& \mu\, {\boldsymbol{\mathsf O}}^{\rm T} \cdot {\bf 1} \cdot{\boldsymbol{\mathsf O}} + \sigma \, {\boldsymbol{\mathsf O}}^{\rm T} \cdot {\boldsymbol{\mathsf G}} \cdot{\boldsymbol{\mathsf O}} \nonumber\\
&=& \mu\, \left(
\begin{array}{ccccc}
N & 0 & 0 & \dots & 0 \\
0 & 0 & 0 & \dots & 0 \\
0 & 0 & 0 & \dots & 0 \\
\vdots & \vdots & \vdots & \ddots & \vdots \\
0 & 0 & 0 & \dots & 0
\end{array}
\right)
+ \sigma \, {\boldsymbol{\mathsf G}}'
\eea
with the matrix
\be
{\boldsymbol{\mathsf G}}'= {\boldsymbol{\mathsf O}}^{\rm T} \cdot {\boldsymbol{\mathsf G}} \cdot{\boldsymbol{\mathsf O}} \, .
\ee
This latter is again a random matrix with elements 
\be
G'_{ij} = \sum_{a,b} ({\boldsymbol{\mathsf O}}^{\rm T})_{ia} G_{ab} ({\boldsymbol{\mathsf O}})_{bj} 
\ee
distributed according to independent Gaussian distribution of zero mean and unit variance.  Indeed, we have that
\be
\langle G'_{ij}\rangle = \sum_{a,b} ({\boldsymbol{\mathsf O}}^{\rm T})_{ia} \langle G_{ab}\rangle ({\boldsymbol{\mathsf O}})_{bj} = 0
\label{Gaussian}
\ee
and
\bea
\langle G'_{ij}G'_{kl}\rangle 
&=& \sum_{a,b,c,d} ({\boldsymbol{\mathsf O}}^{\rm T})_{ia} ({\boldsymbol{\mathsf O}})_{bj} ({\boldsymbol{\mathsf O}}^{\rm T})_{kc} ({\boldsymbol{\mathsf O}})_{dl} \langle G_{ab} G_{cd} \rangle \nonumber\\
&=& \sum_{a} ({\boldsymbol{\mathsf O}}^{\rm T})_{ia} ({\boldsymbol{\mathsf O}})_{ak} \sum_{b} ({\boldsymbol{\mathsf O}}^{\rm T})_{jb}  ({\boldsymbol{\mathsf O}})_{bl} 
= \delta_{ik} \, \delta_{jl} \, .
\eea

If $\mu\ne 0$ and $\sigma\ne 0$, the eigenvalue problem can thus be solved by perturbation theory in the small parameter $\zeta=\sigma/(\mu N)$.  Therefore, in the limit $N\to\infty$, the spectrum of the matrix ${\boldsymbol{\mathsf J}}$ is composed of the eigenvalue
\be\label{lambda-1}
\lambda^{(1)} =\mu\, N + O(N^{0})
\ee
and $N-1$ eigenvalues $\{\lambda^{(m)}\}_{m=2}^{N}$ contained inside the disk of radius $\sigma\sqrt{N}$ in the complex plane $(\Re\,\lambda,\Im\,\lambda)$ with probability one in the limit $N\to\infty$.  The probability density of these latter eigenvalues is similar as for the real Ginibre ensemble in the limit $N\to\infty$ \cite{ginibre_statistical_1965}. In the general case where the matrix elements have non-Gaussian distributions and/or are statistically correlated, the dominant eigenvalue behaves as described by Eq.~(\ref{lambda-1}), but the $N-1$ non-dominant eigenvalues may have different kinds of probability distribution.

\section{General instability criterion}
\label{GIC}

Now using the results of 
section~\ref{Jacobian-matrix} for random matrices with Gaussian independent and identically distributed elements, we conclude that when $\mu N \ge \sigma \sqrt{N}$, which is equivalent to $N \ge (\sigma/\mu)^2$, an isolated eigenvalue will be dominant as in mechanism (i). 
If this condition is satisfied, the instability occurs when $\mu N \ge 1/\tau$, in other words when 
the number of chiral species $N=\NC$ is such that $N_{\rm C}  \ge \max\{1/(\tau \mu), (\sigma/ \mu)^2 \}$ with $\mu>0$.

Instead when $N \le (\sigma/\mu)^2$, the previous eigenvalue is no longer dominant and the instability can only occur due to eigenvalues that are located at the edge of the 
Girko circle \cite{girko_circular_1985}, which is the second mechanism (ii). This edge, which corresponds to eigenvalues with a maximum real part, can be made of either a single real eigenvalue 
or to a pair of conjugated complex valued eigenvalues. 
The instability then occurs when $N \ge 1/(\tau \sigma)^2$, or taken together when
$(\sigma/\mu)^2 \ge N \ge 1/(\tau \sigma)^2$.

We note that the first mechanism (i) holds with $\mu>0$ even for a vanishing root mean square $\sigma=0$, but the second mechanism~(ii) requires that the root mean square $\sigma$ is not equal to zero.

According to this analysis, a non-equilibrium reaction network with sufficiently many chiral species is likely to undergo spontaneous chiral symmetry breaking. We now discuss a specific implementation for a chemical reaction network.

\section{Generalized Frank's model}
\label{GFMSM}


The basic idea is that there exist more chiral than achiral species. Frank's model~\cite{F53} is thus generalized by multiplying the species, especially, the chiral species.  Moreover, the reverse reactions are included in order to possibly satisfy microreversibility.  The reaction network is given by Eqs.~(3)-(4)-(5) of the main text with ${\rm E}_i={\rm D}_i$ and $\bar{\rm E}_i={\rm L}_i$, reading
\bea
&&{\rm A}_a +{\rm D}_i  \rightleftharpoons {\rm D}_j + {\rm D}_k \, , \label{react-ADDD-full} \\
&&{\rm A}_a +{\rm L}_i  \rightleftharpoons {\rm L}_j + {\rm L}_k \, , \label{react-ALLL-full} \\
&&{\rm D}_i + {\rm L}_j   \rightleftharpoons \tilde{\rm A}_b + \tilde{\rm A}_c  \, , \label{react-AADL-full} 
\eea
with $a=1,2,\dots,\NA$; $b,c=1,2,\dots,\NAi$; and $i,j,k=1,2,\dots,\NC$. The net reaction rates are given by
\bea
&& w_{aijk}^{\rm (D)} = k_{+aijk} A_a D_i - k_{-aijk} D_j D_k \qquad\mbox{with}\qquad j\le k\, , \label{rate-ADDD-gen}\\
&& w_{aijk}^{\rm (L)} = k_{+aijk} A_a L_i - k_{-aijk} L_j L_k \qquad\mbox{with}\qquad j\le k \, , \label{rate-ALLL-gen} \\
&& \tilde w_{bcij} = \tilde k_{-bcij} D_i L_j -\tilde k_{+bcij} \tilde A_b \tilde A_c \qquad\mbox{with}\qquad b\le c\, ,\label{rate-AADL-gen}
\eea
where the positive sign in the subscripts of the rate constants refers to the direction of chirality generation. We note that $\tilde k_{\pm bcij}=\tilde k_{\pm bcji}$ because of the mirror symmetry~(\ref{mirror_symm}). The kinetic equations are thus given by
\bea
\dot{A}_d &=& - \sum_{aijk\atop j\le k} \delta_{da}\, w_{aijk}^{\rm (D)} - \sum_{aijk\atop j\le k} \delta_{da}\, w_{aijk}^{\rm (L)} + \frac{1}{\tau} (A_{0d}-A_d) \, , \label{eq-Ad-gen} \\
\dot{D}_m &=& \sum_{aijk\atop j\le k} \nu_{m,ijk}\, w_{aijk}^{\rm (D)}  -\sum_{bcij\atop b\le c} \delta_{mi}\, \tilde w_{bcij} + \frac{1}{\tau} (D_{m0}-D_m) \, , \label{eq-Dm-gen} \\
\dot{L}_m &=&\sum_{aijk\atop j\le k} \nu_{m,ijk}\, w_{aijk}^{\rm (L)} -\sum_{bcij\atop b\le c} \delta_{mj}\, \tilde w_{bcij} +  \frac{1}{\tau} (L_{m0}-L_m) \, , \label{eq-Lm-gen} \\
\dot{\tilde A}_e &=& \sum_{bcij\atop b\le c} (\delta_{eb}+\delta_{ec})\, \tilde w_{bcij} + \frac{1}{\tau} (\tilde A_{0e}-\tilde A_e) \, , \label{eq-Aie-gen} 
\eea
where
\be\label{stoichio}
\nu_{m,ijk} \equiv -\delta_{mi} + \delta_{mj} + \delta_{mk} \, .
\ee

The rate constants can be taken according to log-normal distributions \cite{davidi_birds-eye_2018}. If the rate constants are distributed around some mean values with relatively small root mean squares, the leading behavior can be determined by replacing the rate constants with their mean value. In this respect, we may assume that all the rate constants are equal,
\be\label{Hyp-equal-rates}
\tilde k_{\pm abij} = \tilde k_{\pm} \, , \qquad  k_{\pm aijk} =  k_{\pm} \qquad\left(\forall\ a,b,i,j,k\right) ,
\ee
and the concentrations of achiral species, D-, and L-enantiomers can also be supposed to be equal
\be
A_a=A \, , \qquad D_i = D\, , \qquad L_i=L\, ,\qquad\mbox{and}\qquad \tilde A_a=\tilde A \, , \qquad  \qquad \left(\forall \ a,i \right) .
\ee
Thus, the kinetic equations become
\bea
\dot{A} &=& - \frac{\NC}{\NA}\left[(K_{+} A\, D - K_{-} D^2) + (K_{+} A\, L - K_{-} L^2)\right]+ \frac{1}{\tau} (A_0-A) \, , \label{eq-A-F-NA-NC} \\
\dot{D} &=& (K_{+} A\, D - K_{-} D^2) - (\tilde K_{-} D L- \tilde K_{+}\tilde A^2) + \frac{1}{\tau} (D_{0}-D) \, , \label{eq-D-F-NA-NC} \\
\dot{L} &=& (K_{+} A\, L - K_{-} L^2) - (\tilde K_{-} D L - \tilde K_{+} \tilde A^2) + \frac{1}{\tau} (L_{0}-L) \, , \label{eq-L-F-NA-NC} \\
\dot{\tilde A} &=& 2\, \frac{\NC}{\NAi}\, (\tilde K_{-} D L - \tilde K_{+} \tilde A^2 ) + \frac{1}{\tau} (\tilde A_0-\tilde A) \, , \label{eq-Ai-F-NAi-NC}
\eea
with the effective rate constants:
\be
K_{\pm}\equiv \frac{\NA\NC(\NC+1)}{2}\, k_{\pm} \qquad\mbox{and}\qquad \tilde K_{\pm}\equiv \frac{\NAi(\NAi+1)\NC}{2}\, \tilde k_{\pm} \, .
\ee
As a consequence of these kinetic equations, we have that
\bea
\NA\, \dot{A} +\NC\, \dot{D} +\NC\, \dot{L} +\NAi \dot{\tilde A}&=& \frac{1}{\tau} \big[ (\NA\, A_0 \ + \ \NC\, D_0 \ + \ \NC\, L_0 \ + \ \NAi \, \tilde A_0) \nonumber \\ && - (\NA\, A \ + \ \NC\, D \ + \ \NC\, L \ + \ \NAi \, \tilde A)\big] ,
\eea
implying
\be\label{constraint}
\lim_{t\to\infty} (\NA\, A+\NC\, D+\NC\, L+\NAi \, \tilde A) = \NA\, A_0+\NC\, D_0+\NC\, L_0+\NAi \, \tilde A_0 \, .
\ee

The model is compatible with the existence of equilibrium. Indeed, the detailed balance conditions give the following Guldberg-Waage equilibrium relations,
\be
\frac{D_{\rm eq}}{A_{\rm eq}} = \frac{L_{\rm eq}}{A_{\rm eq}} = \frac{K_+}{K_-} \qquad\mbox{and} \qquad \frac{\tilde A_{\rm eq}}{A_{\rm eq}} =  \frac{K_+}{K_-}  \, \sqrt{\frac{\tilde K_-}{\tilde K_+}} \, .
\ee
Therefore, equilibrium exists for any positive value of the rate constants and two rate constants can independently take arbitrarily small values.

In the fully irreversible regime with $K_-=0$ and $\tilde K_+=0$, we further suppose that the system is only supplied with the achiral species of high free energy: $D_0=L_0=\tilde A_0=0$. Since Eq.~(\ref{constraint}) holds for long enough time, we have that
\be
A = A_0 - \frac{\NC}{\NA} \, (D+L)  - \frac{\NAi}{\NA} \, \tilde A \, .
\ee
Therefore, the kinetic equations reduce to the three following equations:
\bea
\dot{D} &=&  K_{+}  D \left[A_0 - \frac{\NC}{\NA} \, (D+L)  - \frac{\NAi}{\NA} \, \tilde A\right] - \tilde K_{-} D L  - \frac{1}{\tau} \, D \, , \label{eq-D-F-NA-NC-irr3} \\
\dot{L} &=&  K_{+} L \left[A_0 - \frac{\NC}{\NA} \, (D+L)  - \frac{\NAi}{\NA} \, \tilde A\right] - \tilde K_{-} D L - \frac{1}{\tau} \, L \, , \label{eq-L-F-NA-NC-irr3} \\
\dot{\tilde A} &=& 2\, \frac{\NC}{\NAi} \, \tilde K_{-} D L - \frac{1}{\tau} \, \tilde A \, . \label{eq-Ai-F-NAi-NC-irr3}
\eea

Setting
\be\label{Delta}
\Delta\equiv K_+A_0-\frac{1}{\tau} =  \frac{\NC(\NC+1)}{2}\, k_+\, \NA \, A_0 - \frac{1}{\tau} \, ,
\ee
the steady states and their eigenvalues $\{\xi_1,\xi_2,\xi_3\}$ of linear stability are here given by
\bea
D=L=\tilde A=0: && \xi_1=\xi_2=\Delta \, , \ \xi_3=-1/\tau \, , \label{TRS}\\
D=0\, , \ L=\frac{\NA\, \Delta}{\NC\, K_+} \, , \ \tilde A = 0: && \xi_1=-\frac{\NA\, \tilde K_-\, \Delta}{\NC\, K_+} \, , \ \xi_2=-\Delta \, , \nonumber \\ && \xi_3=-1/\tau \, , \label{LHS}\\
D=\frac{\NA\, \Delta}{\NC\, K_+}\, ,\ L=0\, , \ \tilde A = 0: && \xi_1=-\Delta \, , \ \xi_2=-\frac{\NA\, \tilde K_-\, \Delta}{\NC\, K_+}  \, , \nonumber \\ && \xi_3=-1/\tau\, , \label{DHS}\\
D=L=\frac{\NA\, \Delta}{2\NC\, K_++\NA\, \tilde K_-}+O(\Delta^2) \, , \ \tilde A = O(\Delta^2): && \xi_1=\frac{\NA\, \tilde K_-\,  \Delta}{2\NC\, K_++\NA\, \tilde K_-}+O(\Delta^2)
\, , \nonumber\\
&& \xi_2=-  \Delta +O(\Delta^2)\, , \nonumber \\ && \xi_3=-1/\tau +O(\Delta^2) \, .
\label{NTRS}
\eea
In these expressions, the terms $O(\Delta^2)$ are negligible if
\be
\frac{\NA\, \NC \, K_+ \, \tilde K_- \, \tau \, \Delta}{(2\NC\, K_++\NA\, \tilde K_-)^2} \ll 1 \, .
\ee

The behavior is determined by the parameter~(\ref{Delta}). Since concentrations are always non-negative, the only steady state that exists if $\Delta<0$ is the trivial racemic state~(\ref{TRS}), which is an attractor because $\xi_1,\xi_2,\xi_3<0$ in this case.  If $\Delta >0$, three new steady states emerge, which are the L-homochiral attractor~(\ref{LHS}), the D-homochiral attractor~(\ref{DHS}), and the non-trivial racemic state~(\ref{NTRS}).  This latter is unstable since $\xi_1>0$ for this new steady state.
The threshold of instability towards homochirality is thus found at $\Delta=0$. Therefore, spontaneous chiral symmetry breaking happens if the following criterion is satisfied,
\be\label{criterionSM}
k_+ \, \tau \, \NA\, A_0 > \frac{2}{\NC(\NC+1)} \, ,
\ee
which is Eq.~(6) of the main text in the case $\langle k_+\rangle=k_+$ where all the rate constants are equal.

Now, if the rate constants were not all equal as in Eq.~(\ref{Hyp-equal-rates}), but if they were statistically distributed, the analysis carried out here above would provide the mean behavior of the system.  However, the statistical distribution of the rate constants would introduce further effects that should also be analyzed.  In particular, for every steady state, the matrix of linear stability could be decomposed in a similar way as in Eq.~(\ref{J-mu-G}) into a mean value that would be given by Eq.~(\ref{mu}) and fluctuations of root mean square~(\ref{sigma}).  The leading eigenvalue of this random matrix could thus be evaluated as in section~\ref{Jacobian-matrix}, giving an estimation comparable to the eigenvalues $\xi_i$ obtained here above and this for every stable or unstable steady state.

Similar results hold for the other models with either ${\rm E}_m={\rm D}_m$ and $\bar{\rm E}_m={\rm L}_m$ or ${\rm E}_m={\rm L}_m$ and $\bar{\rm E}_m={\rm D}_m$ for 
each enantiomeric pairs $m=1,2,\dots,N_{\rm C}$.

\section{Instability criterion of the trivial racemic state for the generalized Frank model}
\label{Instability_gen_Franck}

Spontaneous chiral symmetry breaking can be investigated by considering the linear stability analysis of any racemic solution with the stationary concentrations ${\bf D}={\bf L}$ where ${\bf D}=\{ D_i\}_{i=1}^{N}$ and ${\bf L}=\{ L_i\}_{i=1}^{N}$.  For this purpose, we introduce the variables
\be\label{EnExcess}
\delta{\bf X} \equiv \frac{1}{2}\left(\delta{\bf L}-\delta{\bf D}\right)  ,
\ee
characterizing infinitesimal deviations with respect to the racemic subspace.  These deviations are ruled by the following set of linear equations:
\be\label{lin-eq}
\frac{d}{dt} \delta{\bf X} = {\boldsymbol{\mathsf M}}\cdot\delta{\bf X}\, ,
\ee
with the matrix
\be
{\boldsymbol{\mathsf M}} = {\boldsymbol{\mathsf J}} - \frac{1}{\tau} \, {\boldsymbol{\mathsf I}}  = \frac{\partial{\bf \dot D}}{\partial{\bf D}} - \frac{\partial{\bf \dot D}}{\partial{\bf L}} \, ,
\ee
since the fundamental chiral symmetry of the kinetic equations implies that
\be
\frac{\partial{\bf \dot D}}{\partial{\bf D}} =\frac{\partial{\bf \dot L }}{\partial{\bf L}}\qquad\mbox{and}\qquad 
\frac{\partial{\bf \dot D}}{\partial{\bf L}} =\frac{\partial{\bf \dot L}}{\partial{\bf D}}\, .
\ee

For the irreversible model we consider, we have the matrix elements
\bea\label{M_mn}
M_{mn} &=&  J_{mn} - \frac{1}{\tau} \, \delta_{mn}  = \frac{\partial \dot D_m}{\partial D_n} - \frac{\partial \dot D_m}{\partial L_n} 
= \sum_{i \atop m\le i} k_{+nmi} \, A + \sum_{i \atop i\le m} k_{+nim} \, A   \\
&+& \tilde k_{-mn} \, D_m  -\delta_{mn} \left( \sum_{ij \atop i\le j} k_{+nij} \, A +\sum_{i} \tilde k_{-mi} \, L_i +\frac{1}{\tau} \right) ,
\nonumber
\eea
where $A$ and $D_i=L_i$ are the concentrations of the stationary racemic solution.  Since the rate constants are supposed to be statistically distributed, this is also the case for the stationary concentrations $\{ D_i\}_{i=1}^{N}$ and $\{ L_i\}_{i=1}^{N}$ and thus for the matrix elements $J_{mn}$.  The statistical distribution of the matrix elements $J_{mn}$ depends on the reaction network and may be complicated, but they could be decomposed as explained in section~\ref{Jacobian-matrix} into a mean value given by Eq.~(\ref{mu}) and fluctuations of root mean square~(\ref{sigma}).

At the trivial racemic fixed point such that $D_i = L_i = \tilde A = 0$ for all species $i$ and $A = A_0$, the elements~(\ref{M_mn}) of the Jacobian matrix associated with the evolution of the enantiomeric excess are evaluated by
\begin{equation}
M_{mn} = J_{mn} - \frac{1}{\tau} \, \delta_{mn}  = A_0 \left( \sum_{i \atop m\le i} k_{+nmi} + \sum_{i \atop i\le m} k_{+nim} \right) - \delta_{mn} \left( A_0 \sum_{ij \atop i\le j} k_{+nij} +\frac{1}{\tau} \right) .
\end{equation}
 Thus the matrix ${\boldsymbol{\mathsf M}}$ can be decomposed into three matrices ${\boldsymbol{\mathsf Q}}$ and ${\boldsymbol{\mathsf R}}$ as
 \begin{equation}
     {\boldsymbol{\mathsf M}} = {\boldsymbol{\mathsf J}} - \frac{1}{\tau} \, {\boldsymbol{\mathsf I}}  = A_0 \, {\boldsymbol{\mathsf Q}} - A_0 \, {\boldsymbol{\mathsf R}} -{\boldsymbol{\mathsf I}}/\tau \, .
     \label{additiveRM}
 \end{equation}
 where ${\boldsymbol{\mathsf I}}$ is the identity, the elements of matrix ${\boldsymbol{\mathsf Q}}$ are given by 
\begin{equation}
    Q_{mn} = \sum_{i \atop m\le i} k_{+nmi} + \sum_{i \atop i\le m} k_{+nim} \, ,
\end{equation}
which are sums of $\NC +1$ random variables of mean $ \langle k_+ \rangle$ and standard deviation $\sigma_{k_+}$ (the element $k_{+nmm}$ occurs twice, once in each sum). Finally, 
${\boldsymbol{\mathsf R}}$ is a diagonal matrix of elements
\begin{equation}
    R_{nn} = \sum_{ij \atop i\le j} k_{+nij} \, .
\end{equation}
According to the central limit theorem, in the large $\NC$ limit, the elements of ${\boldsymbol{\mathsf Q}}$ are distributed following a Gaussian distribution of mean $\mu_Q = \langle k_+ \rangle( \NC + 1)$ and standard deviation $\sigma_Q = \sigma_{k_+}\sqrt{\NC + 1}$.  Moreover, the elements of the matrix ${\boldsymbol{\mathsf R}}$ are also randomly distributed according to a Gaussian of mean $\mu_R = \langle k_+ \rangle \NC(\NC+1)/2$ and standard deviation $\sigma_R = \sigma_{k_+} \sqrt{\NC(\NC+1)/2}$.
Unfortunately, although the spectra of the matrices ${\boldsymbol{\mathsf Q}}$  and ${\boldsymbol{\mathsf R}}$ are known, it is not possible to deduce immediately from these the spectrum of ${\boldsymbol{\mathsf M}}$, because these matrices are not diagonal in the same base.

One can however still use perturbation theory. 
For the matrix ${\boldsymbol{\mathsf Q}}$, we use the same decomposition in terms of a full matrix of ones plus a correction ${\boldsymbol{\mathsf G}}$:
\be
{\boldsymbol{\mathsf Q}} = \mu_Q \, {\bf 1} + \sigma_Q \, {\boldsymbol{\mathsf G}} \, ,
\ee
and we decompose the matrix  ${\boldsymbol{\mathsf R}}$ as 
\be
{\boldsymbol{\mathsf R}} = \mu_R \, {\boldsymbol{\mathsf I}} + {\boldsymbol{\mathsf H}} \, ,
\ee  
where ${\boldsymbol{\mathsf H}}$ is a diagonal matrix with subdominant terms as compared to $\mu_R$ (this follows from the law of large numbers). 
In the end, this means we can decompose ${\boldsymbol{\mathsf M}}$ as
\be
{\boldsymbol{\mathsf M}}=  {\boldsymbol{\mathsf P}} + A_0 \, \sigma_Q \, {\boldsymbol{\mathsf G}} - A_0\,  {\boldsymbol{\mathsf H}} \, .
\ee
where ${\boldsymbol{\mathsf P}}= A_0  \mu_Q {\bf 1} - \left(A_0 \mu_R+\tau^{-1} \right) {\boldsymbol{\mathsf I}}$. The largest eigenvalue of ${\boldsymbol{\mathsf P}}$ is $A_0 (\NC \mu_Q - \mu_R) - \tau^{-1}= A_0 \langle k_+ \rangle \NC(\NC+1)/2 -\tau^{-1}$, and this eigenvalue can be shown to be dominant using the same perturbation calculation as done before.

\begin{figure}[h]
\centerline{\scalebox{0.5}{\includegraphics{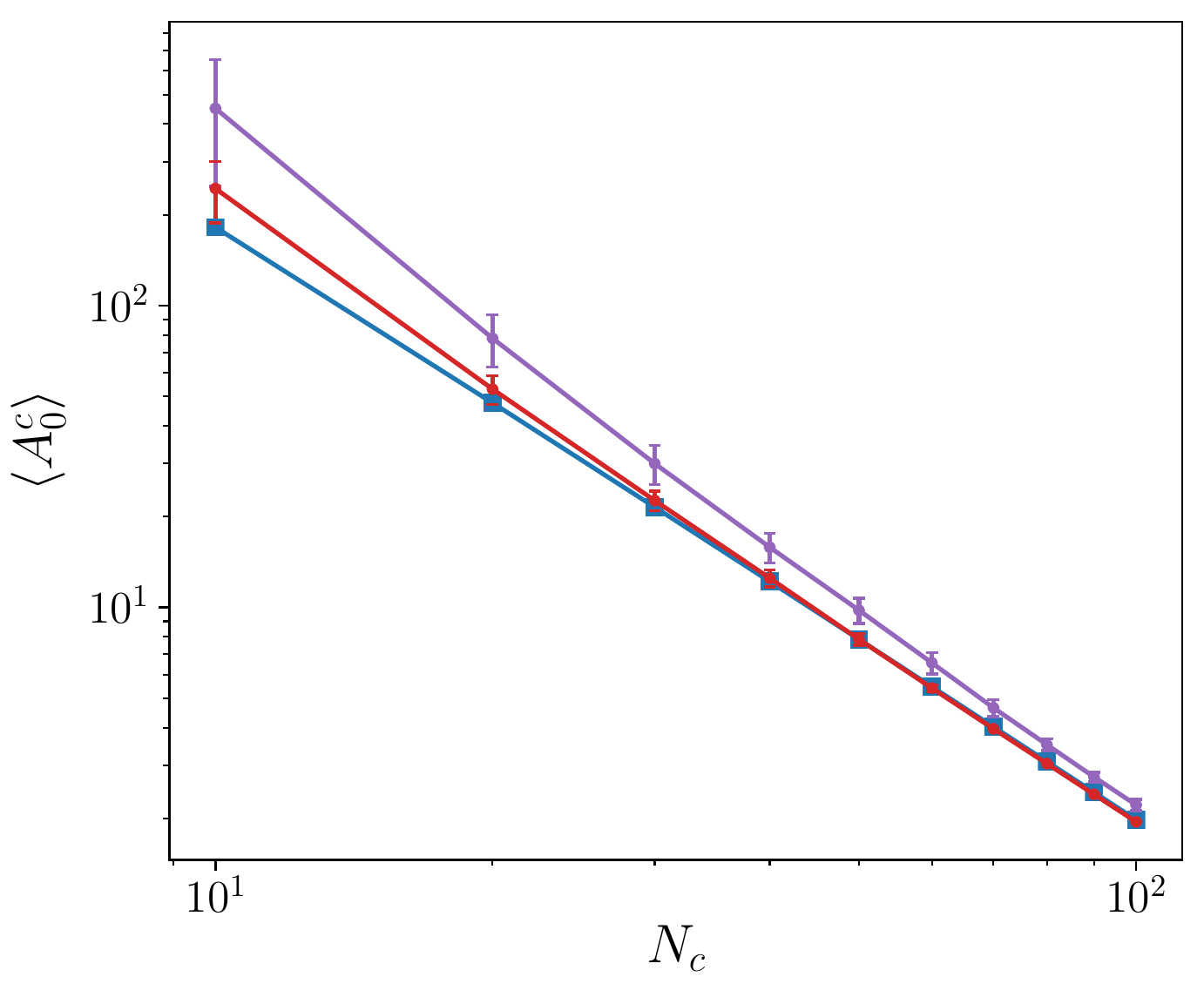}}}
\caption{Comparison between the observed control parameter value $A_0$ at the transition with the theoretical prediction given by  Eq.~(6) of the main text (blue solid line) after averaging over 100 realizations of the rate constants for different standard deviation $\sigma_{k_+}$ and $ \sigma_{\tilde k_-}$ of rate constants : $\sigma_{k_+} = \sigma_{\tilde k_-} = 10^{-3}$ (red), $\sigma_{k_+} = \sigma_{\tilde k_-} = 10^{-2}$ (purple), while $\langle k_+ \rangle = \langle \tilde k_- \rangle = 10^{-4}$. }
\label{verif_criterion}
\end{figure}

For the system to be unstable, this dominant eigenvalue must be positive. Thus, the threshold above which the system is unstable is
\begin{equation}
   \langle k_+ \rangle \, \tau \,  A_0 > \frac{2}{\NC(\NC+1)} \, ,
\label{g_criterion}
\end{equation}
which is Eq.~(\ref{criterionSM}) with the number of achiral species equal to $N_{\rm A} = 1$. Moreover, Eq.~(\ref{criterionSM}) is recovered when all the rate constants are equal.  We note that a deviation from the prediction~(\ref{g_criterion}) is observed when $\sigma_{k_+}$ becomes large compare to $\langle k_+ \rangle$, as depicted in Fig.~\ref{verif_criterion}.

\begin{figure}[h]
\centerline{\scalebox{0.8}{\includegraphics{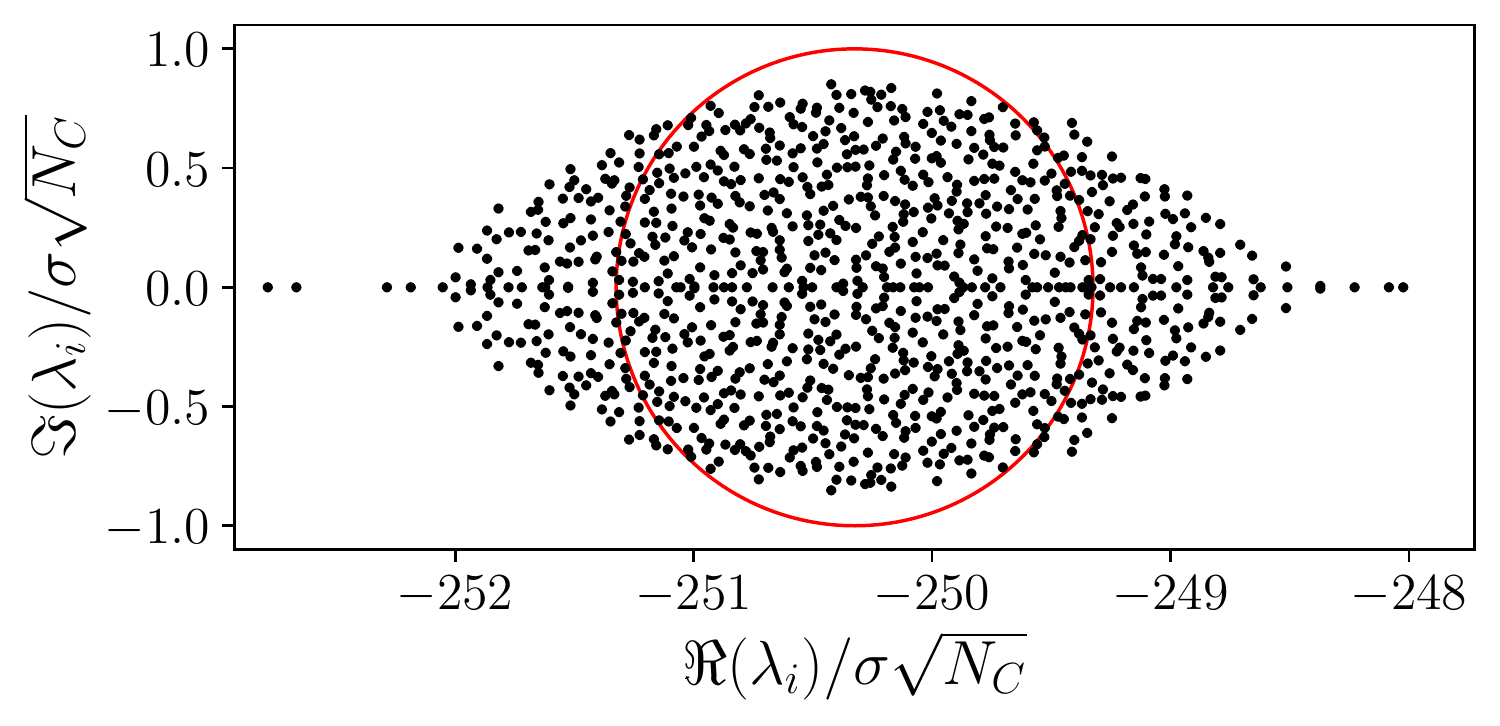}}}
\caption{Close-up of the non-dominant eigenvalues of the Jacobian matrix ${\boldsymbol {\mathsf M}}$ of the generalized Frank model, which do no fill the Ginibre circle.  Here, the matrix ${\boldsymbol {\mathsf M}}$ characterizes a system of $N_{\rm C}=1000$ chiral species, with rate constants distributed according to a log-normal distribution of parameters $\langle k_+ \rangle = 10^{-4}$ and $\sigma_{k_+} = 2 \times 10^{-4}$ and $\tau = 1$. The parameter $A_0 = 25$, is far beyond the instability threshold for the trivial racemic state. The dominant eigenvalue is around 250 and lies outside the field of view for these parameters. The red circle is the unit circle, and the normalization factor $1/\sigma \sqrt{N_{\rm C}}$ is expressed in term of $\sigma =A_0 \sigma_Q$ with $\sigma_Q = \sigma_{k_+}\sqrt{N_{\rm C} +1}$, which is the standard deviation of the matrix ${\boldsymbol {\mathsf Q}}$ elements in the decomposition of the matrix ${\boldsymbol {\mathsf M}}$ in section \ref{Instability_gen_Franck}.}
\label{fig-vp_jacob_zoom}
\end{figure}

One observes that the other eigenvalues of ${\boldsymbol{\mathsf M}}$ do not stay within a 
Girko circle as shown in Fig.~\ref{fig-vp_jacob_zoom}. There is no contradiction since the random matrix ${\boldsymbol{\mathsf M}}$ does not have the same statistics for its diagonal and off-diagonal elements, therefore the assumptions of the Girko theorem do not hold anymore \cite{girko_circular_1985}.

We note that the permutation ${\rm D}_m\leftrightarrow{\rm L}_m$ for some enantiomeric pair $m$ implies that the corresponding enantiomeric excess changes sign, $\delta X_m\to -\delta X_m$.  
However, the eigenvalues of the matrix ${\boldsymbol{\mathsf M}}$ remain unchanged under such transformations.
This can be shown by the following 
calculation. Let us denote the eigenvector ${\boldsymbol{\mathsf u}}$ (with eigenvalue $p$) of the original matrix 
${\boldsymbol{\mathsf M}}$, and the new eigenvector ${\boldsymbol{\mathsf u'}}$ (with eigenvalue $p'$) of the transformed  matrix ${\boldsymbol{\mathsf M'}}$, obtained after such a permutation, 
so that 
\be
\sum_j {\mathsf M}_{ij} \, u_j = p\, u_i    \qquad {\rm and} \qquad \sum_j {\mathsf M}_{ij}' \, u_j' = p' u_i'  \, ,
\ee
Now, the permutation of the enantiomers means that $\delta X_i'=(-1)^{s_i} \delta X_i$ with $s_i=1$ if the $i^{\rm th}$ enantiomers are permuted and $s_i=0$ otherwise.
Using Eq.~(\ref{lin-eq}) for the matrices ${\boldsymbol{\mathsf M}}$ and ${\boldsymbol{\mathsf M'}}$, we obtain $M_{ij}'=(-1)^{s_i+s_j} M_{ij}$. It follows from this that the eigenvectors transform as $u_j'=(-1)^{s_j} u_j$ with no change in the eigenvalues $p'=p$.
Accordingly, all our results hold for the $2^{\NC-1}$ models considered.

\section{Two diffusively coupled compartments}
\label{TDCC}
We consider here two diffusively coupled compartments containing the same chemical network considered before in a well-mixed situation. In addition, we assume the irreversible regime with $N_{\rm A}=\tilde{N}_{\rm \tilde A}=1$ and $N_{\rm C} \gg 1$. The reactions within each compartment read:
\bea
&&{\rm A}_a +{\rm D}_i  \rightarrow {\rm D}_j + {\rm D}_k \, , \label{react-ADDD-full-2} \\
&&{\rm A}_a +{\rm L}_i  \rightarrow {\rm L}_j + {\rm L}_k \, , \label{react-ALLL-full-2} \\
&&{\rm D}_i + {\rm L}_j   \rightarrow 2 \tilde{\rm A}_b  \, , \label{react-AADL-full-2} 
\eea
where $a,b=1$ and $i,j,k=1,\dots,N_{\rm C}$ for species in the first compartment, and $a,b=2$ and $i,j,k=(N_{\rm C}+1),\dots,(2N_{\rm C})$ for species in the second compartment. In addition, there are transfer reactions between compartments for all the species present:
\bea
&& {\rm D}_i \rightleftharpoons {\rm D}_{i+N_{\rm C}}, \\
&& {\rm L}_i \rightleftharpoons {\rm L}_{i+N_{\rm C}}, \\
&& {\rm A}_a \rightleftharpoons {\rm A}_{a+1}, \\
&& \tilde{\rm A}_b \rightleftharpoons \tilde{\rm A}_{b+1},
\eea
which we assume are characterized by the same transition probability $\kappa$ (where the convention that ${\rm E}_i={\rm E}_{i+2N_{\rm C}}$ for the concentrations of enantiomers, ${\rm A}_{a}=A_{a+2}$, and $\tilde{\rm A}_b=\tilde{\rm A}_{b+2}$ is adopted). In the end, the kinetic rate equations of the first reactor are: 
\bea
\dot{A}_d &=& - \sum_{aijk\atop j\le k} \delta_{da}\, w_{aijk}^{\rm (D)} - \sum_{aijk\atop j\le k} \delta_{da}\, w_{aijk}^{\rm (L)} + \frac{1}{\tau} (A_{0d}-A_d) + \kappa (A_{d+1}-A_d) \, , \label{eq-Ad-gen-2} \\
\dot{D}_m &=& \sum_{aijk\atop j\le k} \nu_{m,ijk}\, w_{aijk}^{\rm (D)}  -\sum_{bcij\atop b\le c} \delta_{mi}\, \tilde w_{bcij} + \frac{1}{\tau} (D_{m0}-D_m) + \kappa (D_{m+N}-D_m) \, , \label{eq-Dm-gen-2} \\
\dot{L}_m &=&\sum_{aijk\atop j\le k} \nu_{m,ijk}\, w_{aijk}^{\rm (L)} -\sum_{bcij\atop b\le c} \delta_{mj}\, \tilde w_{bcij} +  \frac{1}{\tau} (L_{m0}-L_m)  + \kappa (L_{m+N}-L_m) \, , \label{eq-Lm-gen-2} \\
\dot{\tilde A}_e &=& \sum_{bcij\atop b\le c} (\delta_{eb}+\delta_{ec})\, \tilde w_{bcij} + \frac{1}{\tau} (\tilde A_{0e}-\tilde A_e) + \kappa (\tilde{A}_{e+1}-\tilde{A}_e), \,  \label{eq-Aie-gen-2} 
\eea
where $d,e=1$; $m=1,\dots,N_{\rm C}$; and $\kappa$ is the diffusive coupling parameter. Similar equations hold for the other reactor, where $d,e=2$ and $m=(N_{\rm C}+1),\dots,(2N_{\rm C})$. From these equations one can proceed by using the enantiomeric excess~(\ref{EnExcess})
which obeys as before the equation: 
\be\label{lin-eq2}
\frac{d}{dt} \delta{\bf X} = {\boldsymbol{\mathsf M}}\cdot\delta{\bf X}\, ,
\ee
with the matrix
\be
{\boldsymbol{\mathsf M}} \equiv \frac{\partial{\bf \dot D}}{\partial{\bf D}} - \frac{\partial{\bf \dot D}}{\partial{\bf L}} \, .
\ee
Now the matrix ${\boldsymbol{\mathsf M}}$ has the following block structure
\begin{equation*}
\label{block-mat}
{\boldsymbol{\mathsf M}} = 
\begin{pmatrix}
{\boldsymbol{\mathsf M}}_1 - \kappa {\boldsymbol{\mathsf I}} & \kappa {\boldsymbol{\mathsf I}} \\
\kappa {\boldsymbol{\mathsf I}} & {\boldsymbol{\mathsf M}}_2 - \kappa {\boldsymbol{\mathsf I}}
\end{pmatrix},
\end{equation*}
where ${\boldsymbol{\mathsf M}}_{1,2}$ represent the Jacobian matrix of compartments $1,2$ respectively and 
${\boldsymbol{\mathsf I}}$ is the identity matrix of same dimension.
In the limit of small $\kappa$, we can treat the effect of diffusion as a small perturbation. This perturbation will introduce a correction of the order of $\kappa$ on the eigenvalues of the uncoupled case ($\kappa=0$). Since the dominant eigenvalues in the uncoupled case are of the order of $\sqrt{N_{\rm C}}$ or $N_{\rm C}$, depending on whether the scenario (i) or (ii) is relevant, this correction should have a small effect on the threshold of instability.

Let us call ${\boldsymbol{\mathsf u}}_1$ (resp. ${\boldsymbol{\mathsf u}}_2$) the eigenvectors of the matrix ${\boldsymbol{\mathsf M}}_1$ (resp. ${\boldsymbol{\mathsf M}}_2$) and the corresponding eigenvalues $p_1$ and $p_2$. A simple calculation provides the eigenvectors ${\boldsymbol{\mathsf u}}$ and the eigenvalues $p$ of the matrix ${\boldsymbol{\mathsf M}}$ as function of ${\boldsymbol{\mathsf u}}_{1,2}$ and $p_{1,2}$. In the case where the dominant eigenvalues of ${\boldsymbol{\mathsf M}}_{1,2}$ are $p_1=p_2=\mu N_{\rm C} \pm O(\sigma)$ for both submatrices, one finds that the dominant contribution to $p$ equals either $\mu N_{\rm C}$ or $\mu N_{\rm C} - 2 \kappa$. The corresponding dominant eigenvectors have respectively uniform components across both compartments: ${\boldsymbol{\mathsf u}}=(1,\dots,1)^{\rm T}$ or opposite components on each compartment: ${\boldsymbol{\mathsf v}}=(1,\dots,1,-1,\dots,-1)^{\rm T}$. The synchronization towards a global homochiral state occurs when the contribution of ${\boldsymbol{\mathsf u}}$ wins over that of ${\boldsymbol{\mathsf v}}$ on long times. Therefore, one then finds that such a synchronization should occur approximately when $\kappa \simeq \sigma/2$.  

Another way to look at the synchronization of the states of the two compartments is to consider 
the evolution of the averaged enantiomeric excess in both compartments, defined by 
\be
\delta \bar{X}_i = \frac{1}{2} \left( \delta X_i + \delta X_{i+N_{\rm C}} \right).
\ee 
The equation of evolution of that quantity is controlled by a matrix $\bar{M}_{mn}$ such that
\be
\bar{M}_{mn}=  \frac{1}{2} \left( M_{mn} + M_{m+N_{\rm C},n} \right).
\ee
In the notation of 
section~\ref{Instability_gen_Franck}, this may be written as
\be
\label{Mbar}
\bar{M}_{mn} = A_0 Q_{mn} - \frac{1}{2} \delta_{mn} (A_0 R_{mm} + 1/\tau) - \frac{1}{2} 
\delta_{m+N_{\rm C},n} (A_0 R_{mm} + 1/\tau),
\ee
where we have used that $Q_{mn}=Q_{m+N_{\rm C},n}$ and $R_{m+N_{\rm C},m+N_{\rm C}}=R_{mm}$. This property holds since the  rate constants take exactly the same values in both compartments because the chemical composition and reactions in the two compartments are exactly the same. Although the concentrations of species take different values in the two compartments, their values do not enter in the stability of the trivial racemic fixed point. 
It follows from Eq.~(\ref{Mbar}), that the eigenvalues of the matrix $\bar{M}$ are exactly the ones we had before in the well-mixed case. From our study of the well-mixed case, we expect that the average enantiomeric excess should undergo an instability when the driving is sufficiently large. If the average enantiomeric excess reaches  extremal values $1$ or $-1$ at long times, then the two compartments must be both homochiral of the same chirality. 

\begin{figure}[h]
\centerline{\scalebox{0.6}{\includegraphics{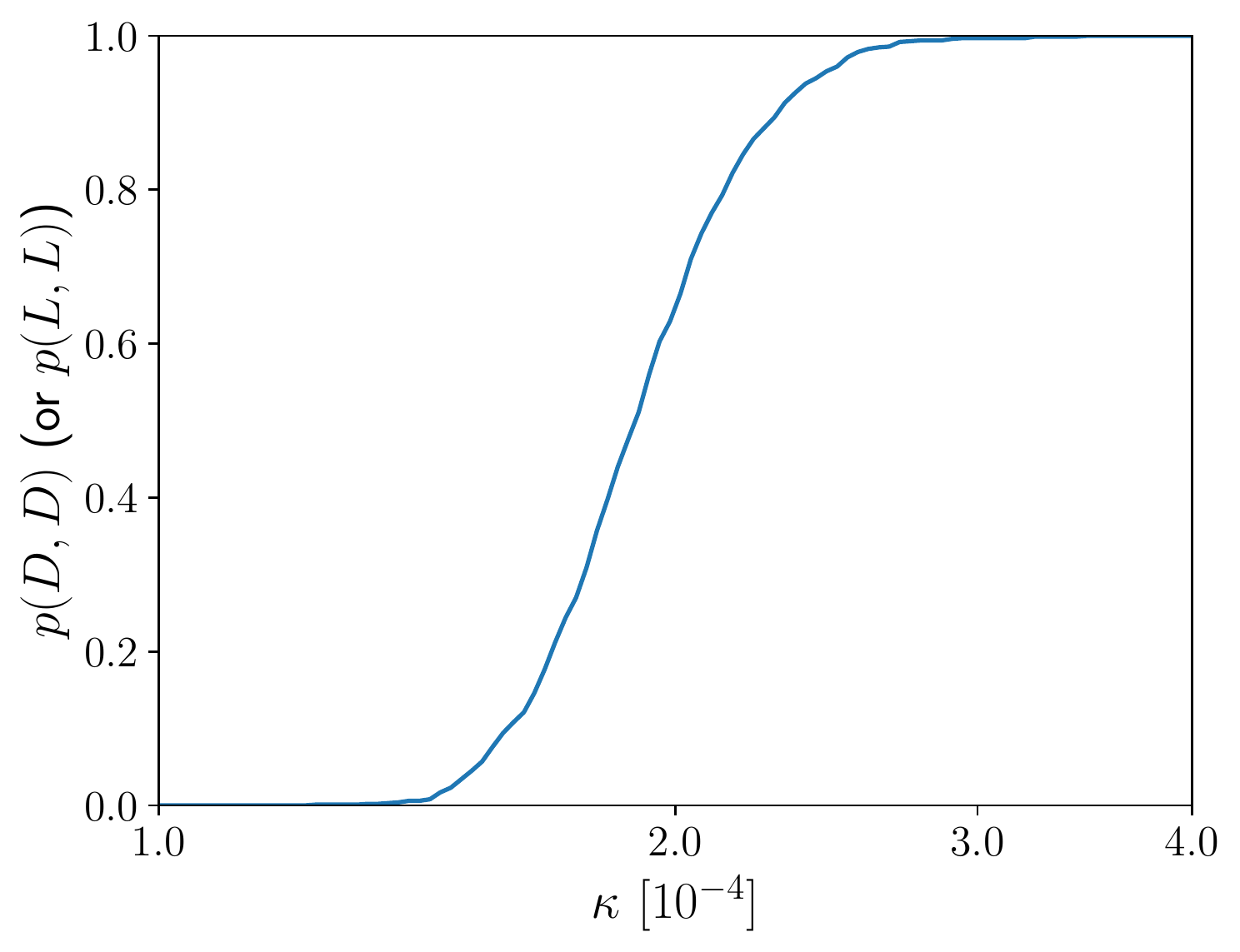}}}
\caption{Probability to have two homochiral compartments of the same chirality on long times (either both are $D$ or both are $L$) as function of the transition probability of transfer of molecules between the two compartments $\kappa$. The initial condition is such that there is an opposite chiral bias in both compartments, so that in the absence of diffusion coupling, the global state will be racemic. For one realization of the rate constants, the parameter $\kappa$ is varied; then this procedure is repeated for different realizations of the rate constants. The curve has been done for a value of the driving force which is above the instability threshold. Simulations were carried out with an initial enantiomeric excess $\epsilon_1 = 10^{-2}$ and D- and L-enantiomers concentrations of all chiral species were initialized at $D_0 = 2 + \epsilon_1$ and $L_0 = 2-\epsilon_1$ in the first compartment and with $\epsilon _2 = 1.5 \times 10^{-2}$ in the second compartment but favoring L-enantiomers. The unactivated achiral specie was initialized at $\tilde A_0 = 0$ and the activated one at $A_0 = 80$, far above the homochirality threshold in each compartment. All the constants $k_{+ijk}$ and $\tilde k_{-ij}$ follow a log-normal distribution of parameters $\mu = -10.02$ and $\sigma = 1.27$ (\textit{i.e.}, corresponding to a log-normal distribution with $\langle k_+ \rangle = \langle \tilde k_- \rangle = 10^{-4}$ and $\sigma_{k_+} = \sigma_{\tilde k_-} = 2\times 10^{-4}$), with $\tilde k_{ij} = \tilde k_{ji}$ to satisfy the mirror symmetry described in Eq.~(\ref{mirror_symm}). The number of chiral species was set up to $\NC = 20$.}
\label{fig-2patches}
\end{figure}

Using numerical simulations, we have confirmed this scenario. Firstly, when diffusion is weak for $\kappa \to 0$, we recover the previous scenario for a transition to homochirality, separately holding in each compartment
As $\kappa$ increases, so does the coupling between the two compartments. If there is a small bias present which is the same in the two compartments (L for instance), then one ends up with a homochiral state which is L in that case. The interesting case is therefore when the two compartments are given an opposite small bias initially. Then, as shown in Fig. \ref{fig-2patches}, we find that as $\kappa$ increases, we go from a global racemic state at small values of $\kappa$ towards a global homochiral state when the coupling is sufficiently strong. The threshold of instability is found not to be significantly changed as compared to the well-mixed case in agreement with the theoretical argument given above.
In addition, the threshold where the transition occurs is indeed of the order of $\sigma=2 \times 10^{-4}$.

\newpage

\newpage

\newpage

\newpage

\newpage

\newpage

\newpage

\end{document}